\documentclass[aip,jap,amsmath,amssymb,twocolumn,10pt]{revtex4-1}
\usepackage{amsmath}
\usepackage{cases}
\usepackage{amssymb}

\usepackage{epstopdf}
\usepackage{graphicx}
\usepackage{color}
\usepackage{multirow}

\begin{document}
\title{First principles calculations of the interface properties of amorphous-$\rm Al_2O_3$/$\rm MoS_2$ under non-strain and biaxial strain conditions}
\author{Li-Bin Shi}
\email{slb0813@126.com; shilibin@bhu.edu}
\author{Ming-Biao Li}
\author{Xiao-Ming Xiu}
\author{Xu-Yang Liu}
\author{Kai-Cheng Zhang}
\author{Yu-Hui Liu}
\author{Chun-Ran Li}
\author{Hai-Kuan Dong}
\affiliation
{School of Mathematics and Physics, Bohai University, Liaoning Jinzhou 121013, China}
\date{\today}
\begin{abstract}
$\rm Al_2O_3$ is a potential dielectric material for metal-oxide-semiconductor (MOS) devices. $\rm Al_2O_3$ films deposited on semiconductors usually exhibit amorphous due to lattice mismatch. Compared to two-dimensional graphene, $\rm MoS_2$ is a typical semiconductor, therefore, it has more extensive application. The amorphous-$\rm Al_2O_3$/$\rm MoS_2$ (a-$\rm Al_2O_3$/$\rm MoS_2$) interface has attracted people's attention because of its unique properties. In this paper, the interface behaviors of a-$\rm Al_2O_3$/$\rm MoS_2$ under non-strain and biaxial strain are investigated by first principles calculations based on density functional theory (DFT). First of all, the generation process of a-$\rm Al_2O_3$ sample is described, which is calculated by molecular dynamics and geometric optimization. Then, we introduce the band alignment method, and calculate band offset of a-$\rm Al_2O_3$/$\rm MoS_2$ interface. It is found that the valence band offset (VBO) and conduction band offset (CBO) change with the number of $\rm MoS_2$ layers. The dependence of leakage current on the band offset is also illustrated. At last, the band structure of monolayer $\rm MoS_2$ under biaxial strain is discussed. The biaxial strain is set in the range from -6\% to 6\% with the interval of 2\%. Impact of the biaxial strain on the band alignment is investigated.
\end{abstract}
\maketitle
\section{Introduction}
Since integrated circuit was born in 1958, the microelectronic technology has made rapid development. Moore's law predicts that the number of metal-oxide-semiconductor-field-effect-transistors (MOSFETs) on a chip doubles every 18 months.\cite{moore1998cramming} For a long time, the gate oxide of integrated circuit has been using silicon dioxide. In order to further reduce the feature size of integrated circuits, we must abandon silicon dioxide, and select the materials with higher relative dielectric constant (high-$\kappa$).\cite{he2013interface,he2011integrations} If high-$\kappa$ materials are used to replace conventional silicon dioxide as the gate dielectric material, the physical thickness of the gate dielectric layer can be increased, so that the gate leakage current can be greatly suppressed. In order to develop field effect transistors with lower power consumption and higher efficiency, people have done a lot of research on high-$\kappa$ materials, including $\rm Ta_2O_5$,\cite{fang2012preliminary} $\rm Si_3N_4$,\cite{park1996characteristics} $\rm Gd_2O_3$, \cite{hong1999epitaxial} $\rm La_2O_3$, \cite{chiu2005electrical} $\rm Y_2O_3$,\cite{wu2015single} $\rm Sc_2O_3$,\cite{cai2007conduction} $\rm Ga_2O_3$,\cite{hsieh2008coaxial} $\rm Lu_2O_3$, \cite{xiong2009electronic} $\rm Al_2O_3$,\cite{guo2013nitrogen,choi2013impact,hoex2008on,werner2011electronic,lin2013chemical,suh2013al2o3/tio2} $\rm LaAlO_3$, \cite{liu2013interfacial} $\rm SrTiO_3$,\cite{choi2007dual} $\rm LaLuO_3$,\cite{olyaei2012low-frequency} $\rm ZrO_2$,\cite{Zheng2007First} $\rm HfO_2$, \cite{kang2003first-principles} Hf silicate,\cite{xiong2007defect} and Zr silicate.\cite{cota2013solution} $\rm Al_2O_3$ among them has become a promising candidates due to its larger band gap (5$\sim$9 eV) and moderate dielectric constant (8$\sim$10).
\par
$\rm Al_2O_3$ thin films deposited on semiconductor substrates usually display amorphous due to lattice mismatch between $\rm Al_2O_3$ and semiconductors. \cite{Costina2001Band,Liu2013Electrical,Hu2014Optical,Chagarov2008Generation,Guti2002Molecular,lamparter1997structure} Amorphous-$\rm Al_2O_3$ (a-$\rm Al_2O_3$) has a complex structure, and knowledge of its microstructure plays an important role in the analysis of the oxidation and passivation details of aluminum. A-$\rm Al_2O_3$/semiconductor interface is expected to be superior to crystalline-$\rm Al_2O_3$/semiconductor interface due to lower interface defect density. At present, a large number of theoretical and experimental studies have been done on crystalline $\rm Al_2O_3$,\cite{guo2013nitrogen,choi2013impact,hoex2008on,werner2011electronic,lin2013chemical,suh2013al2o3/tio2} which mainly include the crystal structure, defect formation and electronic structure. There is not much research on a-$\rm Al_2O_3$ as a dielectric due to the complex microscopic structure. In particular, there is less theoretical research on the formation mechanism of a-$\rm Al_2O_3$.
\par
Since the discovery of graphene in 2004, two-dimensional materials have attracted people's attention.\cite{Novoselov2004Electric,Nguyen2016Effect,Lanzillo2016Band} Low dimensional materials are interesting not only because they can provide access to novel physical phenomena, but also because their unique electrical, optical and mechanical properties make them the focus of attention.\cite{Chang2015Modeling,Chang2013Atomistic,Banerjee2016Performance,Nishiguchi2015Observing,Kang2013Band} $\rm MoS_2$ crystal is composed of Mo atomic layer sandwiched between two layers of S, forming a triangular prismatic arrangement.\cite{Han2010Band, Das2012High} The Mo-S bonding is strong covalent, but the coupling between $\rm MoS_2$ monolayer is weak van der Waals interactions. Because monolayer $\rm MoS_2$ is a typical semiconductor, it is considered as promising candidates for nanoelectronics applications. However, one of the major limiting factors for low dimensional materials is the interface. At present, people have done some research on $\rm Al_2O_3$ and $\rm MoS_2$.\cite{Singh2015Al2O3,Son2015Improved} However, the $\rm Al_2O_3$/$\rm MoS_2$ interface properties are paid little attention. We have not found a detailed study of the a-$\rm Al_2O_3$/$\rm MoS_2$ interface under biaxial
strain.
\par
In this paper, we have done three aspects of studies. First, we investigate generation of a-$\rm Al_2O_3$ by first principles molecular dynamics simulations. Then, we analyze band alignment between a-$\rm Al_2O_3$ and $\rm MoS_2$. Impact of the $\rm MoS_2$ layer thickness on band offset is discussed. At last, we investigate effects of the biaxial strain on a-$\rm Al_2O_3$/$\rm MoS_2$ interface. Impact of the band offset on leakage current is discussed.

\section{Computational details}
\subsection{Generation of amorphous-$\rm Al_2O_3$ sample}
Amorphous-$\rm Al_2O_3$ is generated by melting and quenching technique. Molecular dynamics with NVT ensemble and geometric optimization based on density functional theory (DFT) are performed in the calculation. Generalized gradient approximation (GGA) of Perdew, Burke, and Ernzerhof (PBE) is choose as  exchange correlation functional during molecular dynamics simulation.\cite{perdew1996generalized} All calculations are carried out in CASTEP code based on the total-energy pseudopotential method.\cite{segall2002first-principles}
\par
Figure 1 describes the specific generation process of a-$\rm Al_2O_3$ sample. It is prepared by starting with a monoclinic crystal corresponding to $\theta$-$\rm Al_2O_3$ at the density of 3.61 $\rm g/cm^3$, which is presented in Fig. 1 (a). Its lattice constants are a=11.80\r{A}, b=2.91\r{A}, and c=5.62\r{A}. In order to perform molecular dynamics simulations, we cut $\theta$-$\rm Al_2O_3$ along (100) direction, and construct a orthogonal $\rm Al_2O_3$ supercell by $3\times2\times1$ extension. The supercell consists of 72 O and 48 Al atoms with lattice constants of a=8.73\r{A}, b=11.24\r{A}, and c=11.50\r{A}, which is shown in Fig. 1(b). High temperature annealing at low density provides very good oxide mixing and completely removes the original ordered geometry. The low density sample is formed by rescaling the supercell box size along every direction. In order to ensure the full mixing between the oxides, we construct the low density $\rm Al_2O_3$ sample with value of 0.62 $\rm g/cm^3$ by expanding the supercell box size 1.8 times. Fig. 1 (c) presents the sample prepared by annealing at 5000 K for 1 ps with time step of 1 fs. In order to obtain experimental results, we must gradually increase the sample density. We first increase the sample density to 1.32 $\rm g/cm^3$ by rescaling the supercell box size from 1.8 times to 1.4 times, subsequent sample is annealed at 5000 K for 0.5 ps with time step of 1 fs. Fig. 1 (d) shows the sample after annealing at 5000 K. Then, the sample density is increased to 2.37 $\rm g/cm^3$ by rescaling the supercell box size from 1.4 times to 1.15 times. Fig. 1 (e) shows the sample annealed at 5000 K for 0.5 ps with time step of 1 fs. Finally, we increase the sample density to 3.4 $\rm g/cm^3$ by rescaling the supercell box size, which is in agreement with experimental values of 3.05 $\rm g/cm^3$ $\thicksim$ 3.40 $\rm g/cm^3$ for a-$\rm Al_2O_3$ sample \cite{Lee1995Thermal,Oka1979Structural}. The sample is annealed at 5000 K for 0.5 ps with time step of 1 fs, and quickly cooled up to 10 K. Fig. 1 (f) presents the cooled sample. In order to eliminate the internal stress, we do geometric optimization on samples, which is shown in Fig. 1 (g).
\par
The electronic structure analysis for the annealed and relaxed a-$\rm Al_2O_3$ indicates a band gap of 3.66 eV, which is in agreement with a previous DFT band gap of 3.80 eV.\cite{Chagarov2008Generation} It is well known that the GGA underestimate band gaps of semiconductors or oxides, which leads to inaccurate calculations of band alignments. The hybrid Heyd-Scuseria-Ernzerhof functional (HSE) combines screened Hartree Fock exchange with the GGA-PBE \cite{heyd2003hybrid,heyd2004efficient,paier2006erratum,heyd2005energy}, which can give an accurate description on band gaps of semiconductors or oxides \cite{lyons2011the,dewalle2013defects,choi2013native}. Table 1 presents the band gaps of some semiconductors and oxides calculated by different exchange correlation functionals. The results indicate that the values calculated by HSE functional are close to the experiments. The a-$\rm Al_2O_3$ band gap is increased to 5.26 eV by applying the HSE functional. The experimental values of a-$\rm Al_2O_3$ band gap is in the range from 3.2 eV to 6.7 eV,\cite{Costina2001Band,Liu2012Interfacial} which is closely related to the film growth technology. By comparison, we consider that the band gap value calculated by HSE functional is reasonable.
\par
Table 2 presents averaged bond lengths of O-O, Al-O and Al-Al for our a-$\rm Al_2O_3$ sample, which is compared with previous simulations and experiments. Our results are found to be very close to previous simulations and experiments, which indicates that our a-$\rm Al_2O_3$ sample is close to the actual situation.
\begin{table}
\linespread{1.3}
\centering
  \caption{Band gaps of some semiconductors and oxides, electron volt (eV) as a unit.}
  \label{tb1}
  \footnotesize
  \begin{tabular}{ccccc}
    \hline
    \hline
    Material                 &$\rm LDA$    &$\rm GGA$       &$\rm HSE$         &$\rm Experiment$ \\
    \hline
    $\rm Si$                   &0.64         &0.73           &1.19            &1.12  \cite{afanas2006electron} \\
    $\rm GaAs$                 &0.65         &0.89           &1.39            &1.42  \cite{windhorn1982electron} \\
    monolayer $\rm MoS_2$      &1.84         &1.63           &2.11            &1.90  \cite{Mak2010Atomically} \\
    bulk $\rm MoS_2$           &0.60         &0.98           &1.46            &1.30  \cite{Mak2010Atomically} \\
    $\rm ZrO_2$                &3.79         &3.79           &5.39            &5.83  \cite{french1994experimental} \\
    $\rm HfO_2$                &4.52         &4.67           &6.17            &5.80  \cite{he2007optical} \\
    $\rm Y_2O_3$               &4.17         &4.27           &5.68            &6.00  \cite{ohta2004photoelectron} \\
    $\rm La_2O_3$              &3.49         &3.54           &5.08            &5.18  \cite{qiya2014band} \\
    $a-\rm Al_2O_3$            &3.47         &3.66           &5.26            &3.2$\thicksim$6.7\cite{Costina2001Band,Liu2012Interfacial} \\
    $\alpha-\rm Al_2O_3$       &5.87         &5.90           &8.65            &8.80  \cite{1990electronic} \\
    $\kappa-\rm Al_2O_3$       &4.95         &4.99           &6.68            &   \\
    $\theta-\rm Al_2O_3$       &4.54         &4.59           &6.28            &   \\
    \hline
    \hline
\end{tabular}
\end{table}
\begin{table}
\linespread{1.3}
\centering
  \caption{Averaged bond lengths of O-O, Al-O and Al-Al for  a-$\rm Al_2O_3$ sample versus previous simulation and experimental data. Angstrom (\r{A}) as a unit.}
  \label{tb1}
  \footnotesize
  \begin{tabular}{cccc}
    \hline
    \hline
    Bond length            &Our Sample     &Previous simulation\cite{Guti2002Molecular}    &Experiment \cite{lamparter1997structure} \\
    \hline
    $\rm Al-O$             &1.82           &1.76                       &1.8\\
    $\rm O-O$              &2.70           &2.75                       &2.8\\
    $\rm Al-Al$            &3.05           &3.12                       &3.2\\
    \hline
    \hline
\end{tabular}
\end{table}
\begin{figure*}[htp]
\centering
\includegraphics[width=16cm]{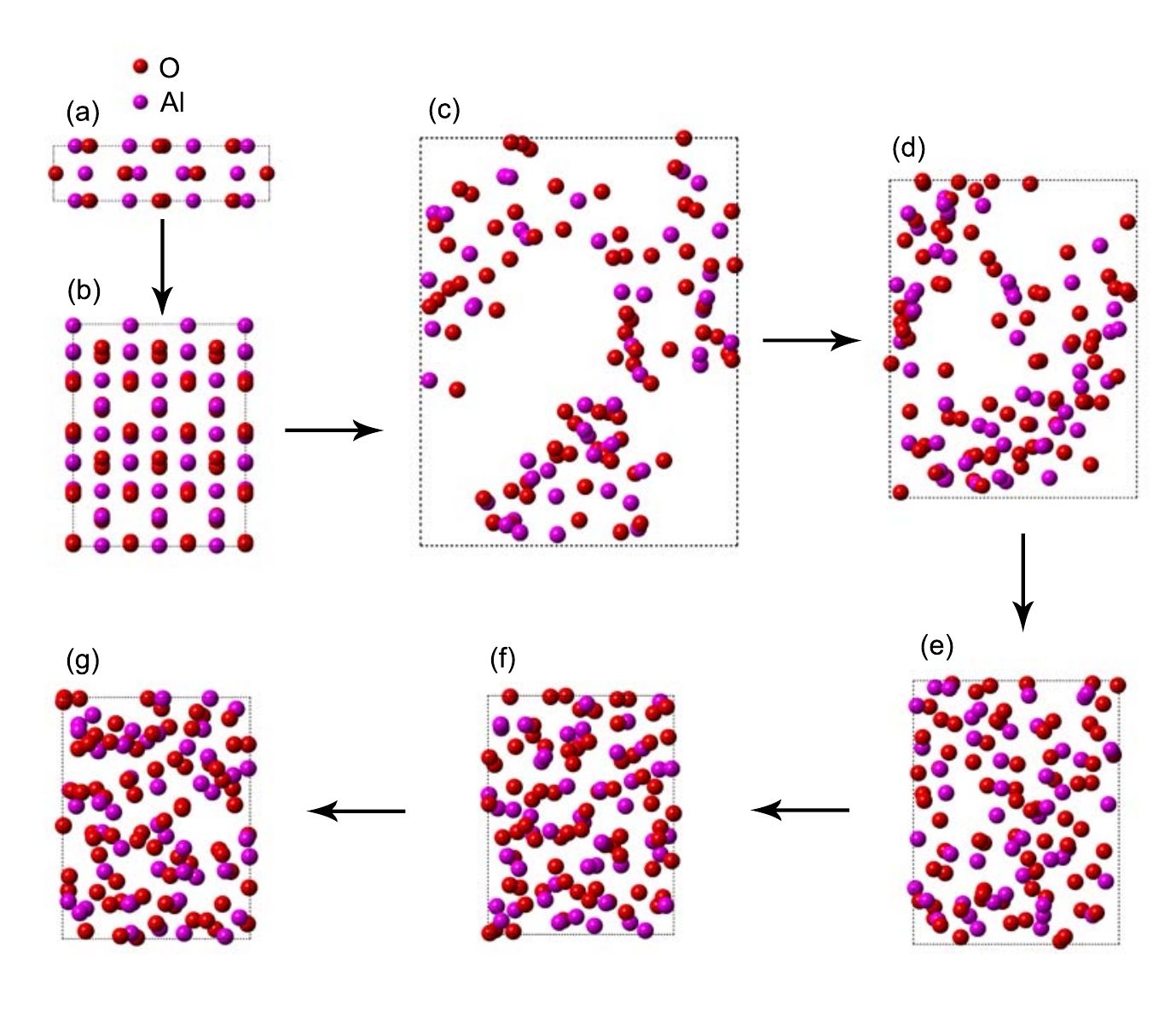}
\caption{Generation process of a-$\rm Al_2O_3$ sample. (a) $\theta$-$\rm Al_2O_3$ at the density of 3.61 $\rm g/cm^3$, (b) $\theta$-$\rm Al_2O_3$ supercell including 72 O and 48 Al atoms, (c) after annealing at 5000 K with low density of 0.62 $\rm g/cm^3$, (d) after annealing at 5000 K with low density of 1.32 $\rm g/cm^3$, (e) after annealing at 5000 K with low density of 2.37 $\rm g/cm^3$, (f) after annealing at 5000 K with low density of 3.40 $\rm g/cm^3$, (g) after geometric optimization.}
\label{Fig:1}
\end{figure*}
\subsection{Band alignment methods}
One of the most important features on oxide/semiconductor interfaces is band offset, i.e., the relative energy level positions on both sides of the interface. The valence band offset (VBO) can be defined as the difference between positions of valence band maximum (VBM), which can be obtained by calculating the band structure and average electrostatic potential (AEP). In the potential line up method, the VBO is usually split into two terms. \cite{silvestri2013first,d2012first}
\begin{equation}
\rm VBO=\Delta E_v+\Delta V
\end{equation}
The first contribution of $\rm \Delta E_v$ corresponds to alignment of VBM for bulk band structure term of oxide and semiconductor. The second term of $\rm \Delta V_s$ corresponds to the macroscopic AEP alignment, which can be determined by two methods. One method is to determine the macroscopic AEP by constructing the oxide/semiconductor interface. The other is to determine the macroscopic AEP by constructing oxide and semiconductor surfaces. $\rm SrTiO_3$/$\rm TiO_2$ interface has been extensively investigated because the in-plane lattice mismatch between $\rm SrTiO_3$ and $\rm TiO_2$ along (001) direction is less than 3\%. In order to verify the validity of two methods, we calculate the VBO by constructing $\rm SrTiO_3$ and $\rm TiO_2$ surfaces as well as $\rm SrTiO_3$/$\rm TiO_2$ interface. It is found that the surface and interface calculations can give similar results, which is also found in previous investigation of $\rm Al_2O_3$/III-V interface \cite{weber2011native}. The VBO calculated by constructing material surface is similar to that calculated by calculating the work function of the material, which has been widely used to study the band offset. \cite{Liu2007First,Kang2013Band}
\par
In this paper, we align the macroscopic AEP by constructing a-$\rm Al_2O_3$ and $\rm MoS_2$ surfaces instead of a-$\rm Al_2O_3$/$\rm MoS_2$ interface. The reasons are as follows. (a) There is a large difference in the atomic arrangement of a-$\rm Al_2O_3$ and $\rm MoS_2$. The atoms near interface can only reach equilibrium state through larger migration. The atomic migration causes the AEP distortion, which lead to a inaccurate alignment of macroscopic AEP. However, this problem can be avoided by a-$\rm Al_2O_3$ and $\rm MoS_2$ surfaces. (b) The interface structure will produce a larger cell, which takes more time to calculate. Especially for the nonlocal HSE calculation, the cell size will have a great influence on the calculation time.
\par
A-$\rm Al_2O_3$ surface contains 144 O and 96 Al atoms. In order to avoid the interaction between the top and bottom layer, a vacuum region of 16 \r{A} is built to separate them. $\rm MoS_2$ surface containing layer number from 1 to 6 is built, which has a vacuum region of 16 \r{A}. The monolayer $\rm MoS_2$ surface contains 18 S and 9 Mo atoms. The cores for all atoms are represented by norm-conserving pseudopotential, while the valences states are expanded in a plane-wave basis set with 600 eV.
\section{Results and discussion}
\subsection{Band alignment of a-$\rm Al_2O_3$/$\rm MoS_2$ interface}
In order to elucidate band gap shift, we calculate the band structures for bulk and different thickness layers of $\rm MoS_2$ by GGA-PBE and HSE functionals as shown in Fig. 2. The band structures suggest that bulk $\rm MoS_2$ is a semiconductor with indirect band gap, the VBM at G-point, and the conduction band minimum (CBM) between K-point and G-point. This calculation is consistent with previous investigation.\cite{Han2010Band} Calculated band gap is 0.98 eV for GGA-PBE and 1.46 eV for HSE. It is found that HSE band gap is close to experimental value of 1.30 eV.\cite{Mak2010Atomically} For 4-layer $\rm MoS_2$, the band gap is increased to 1.06 eV for GGA-PBE and 1.54 eV for HSE. We are surprised to find that the CBM shifts to K-point as $\rm MoS_2$ changes from bulk structure to 4-layer $\rm MoS_2$. However, the positions of VBM can not be found to change significantly. It is noted that the band gap increases with the decrease of the number of $\rm MoS_2$ layers. The GGA-PBE and HSE band gaps are increase to 1.09 eV and 1.58 eV for 3-layer $\rm MoS_2$, and 1.19 eV and 1.69 eV for 2-layer $\rm MoS_2$. Interestingly, the position of VBM shifts from G-point to K-point as $\rm MoS_2$ changes from 2-layer $\rm MoS_2$ to 1-layer $\rm MoS_2$ (monolayer $\rm MoS_2$). Therefore, it changes from indirect gap semiconductor to direct band gap semiconductor. The GGA-PBE and HSE band gaps are increase to 1.63 eV and 2.11 eV for 1-layer $\rm MoS_2$. The experimental value of band gap for monolayer $\rm MoS_2$ is 1.90 eV.\cite{Mak2010Atomically} Comparing to the experimental values, it is found that the band gap is underestimated by GGA-PBE while it is slightly overestimated by HSE.
\begin{figure*}[htp]
\centering
\includegraphics[width=16 cm]{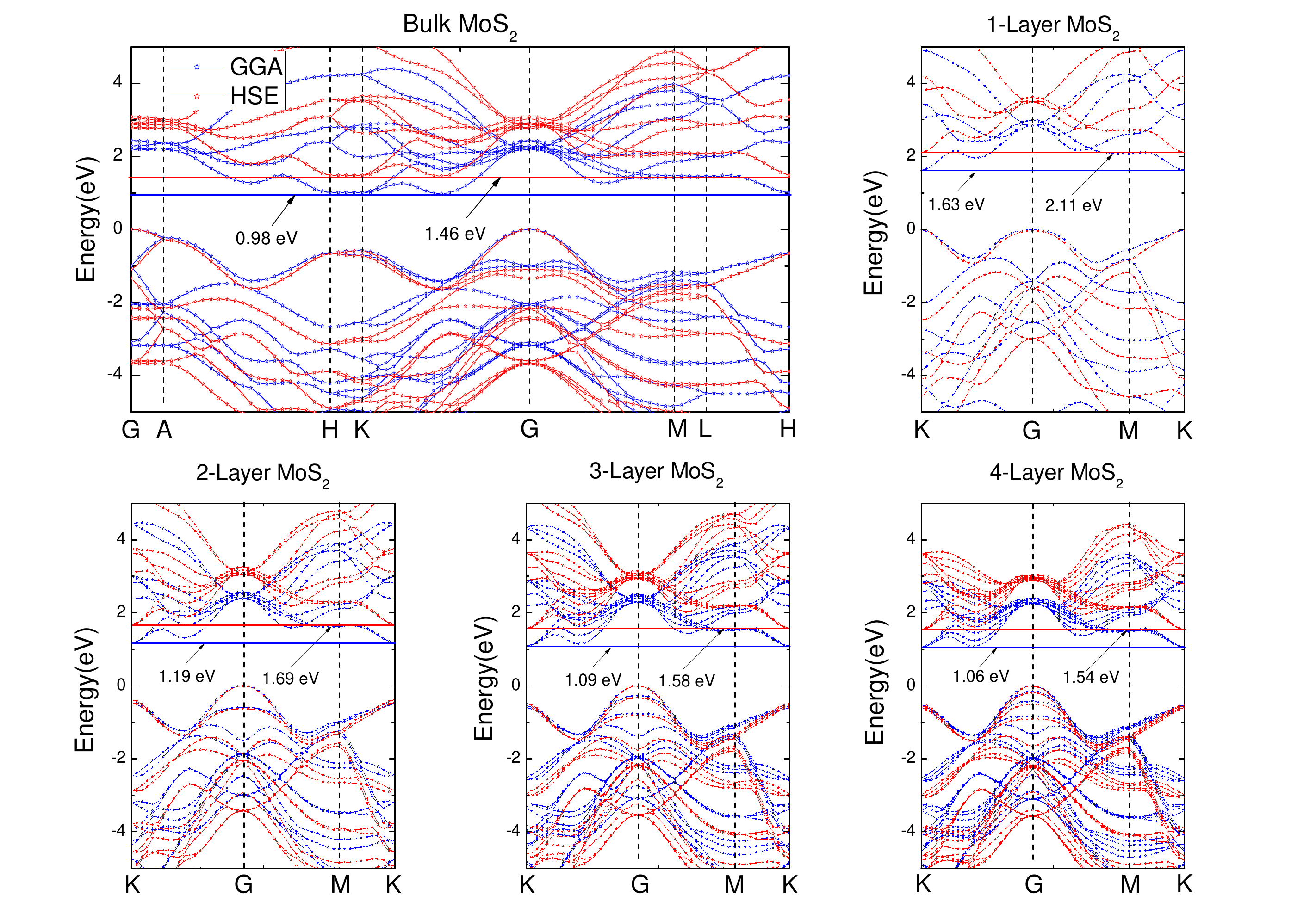}
\caption{Band structures for bulk and different thickness layer of $\rm MoS_2$. The blue and red point connections represent GGA-PBE and HSE results. The VBM is set to zero in order to check the band gap. The blue and red horizontal solid line in each panel indicates the CBM calculated by GGA-PBE and HSE functionals. The arrows indicate the band gap values. Special points in Brillouin zone are set as G (0 0 0); A (0 0 0.50); H (-0.33 0.67 0.50); K (-0.33 0.67 0); M (0 0.50 0); L (0 0.50 0.50).}
\label{Fig:2}
\end{figure*}
\par
Figure 3 shows a-$\rm Al_2O_3$ and $\rm MoS_2$ surfaces as well as planar and macroscopic AEP. In order to align the AEP, the vacuum level is scaled to zero. Macroscopic AEP is the averaged value of planar AEP, which is represented by a red point connection in Fig. 03. For $\rm MoS_2$ surface, the planar AEP in the atomic region exhibits a periodic oscillation, while it remains constant in the vacuum region.\cite{Li2016Investigation} It is found that the shape of planar AEP do not change as the number of the $\rm MoS_2$ layers changes from 1 to 6. For a-$\rm Al_2O_3$ surface, the planar AEP in the atomic region has no obvious periodicity. The value of macroscopic AEP for $\rm MoS_2$ surface is -13.06 eV while its values is -18.31 eV for a-$\rm Al_2O_3$.
\begin{figure*}
\centering
\includegraphics[width=12 cm]{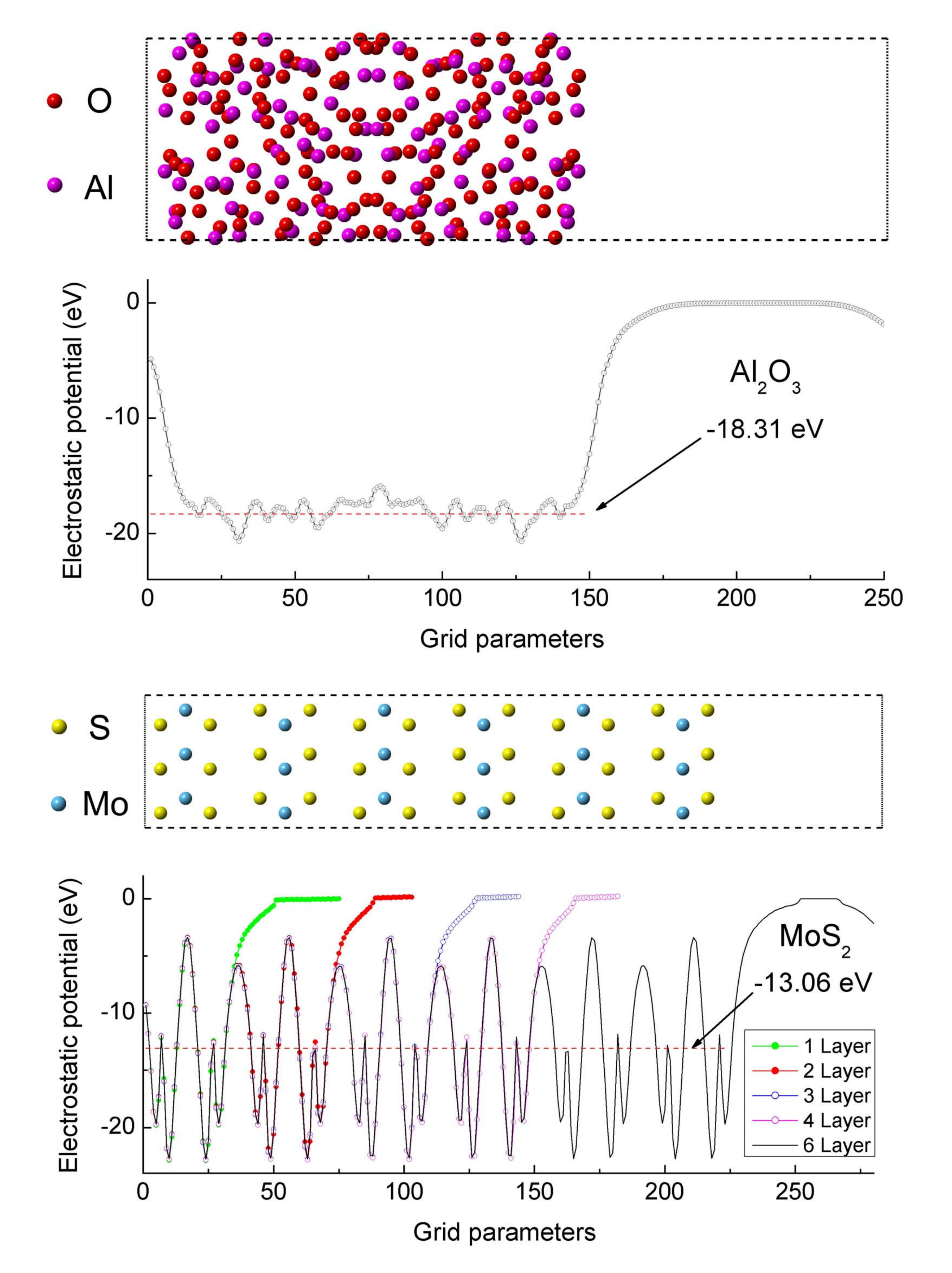}
\caption{a-$\rm Al_2O_3$ and $\rm MoS_2$ surfaces as well as planar and macroscopic AEP.}
\label{Fig:3}
\end{figure*}

\begin{figure}[htp]
\centering
\includegraphics[width=8 cm]{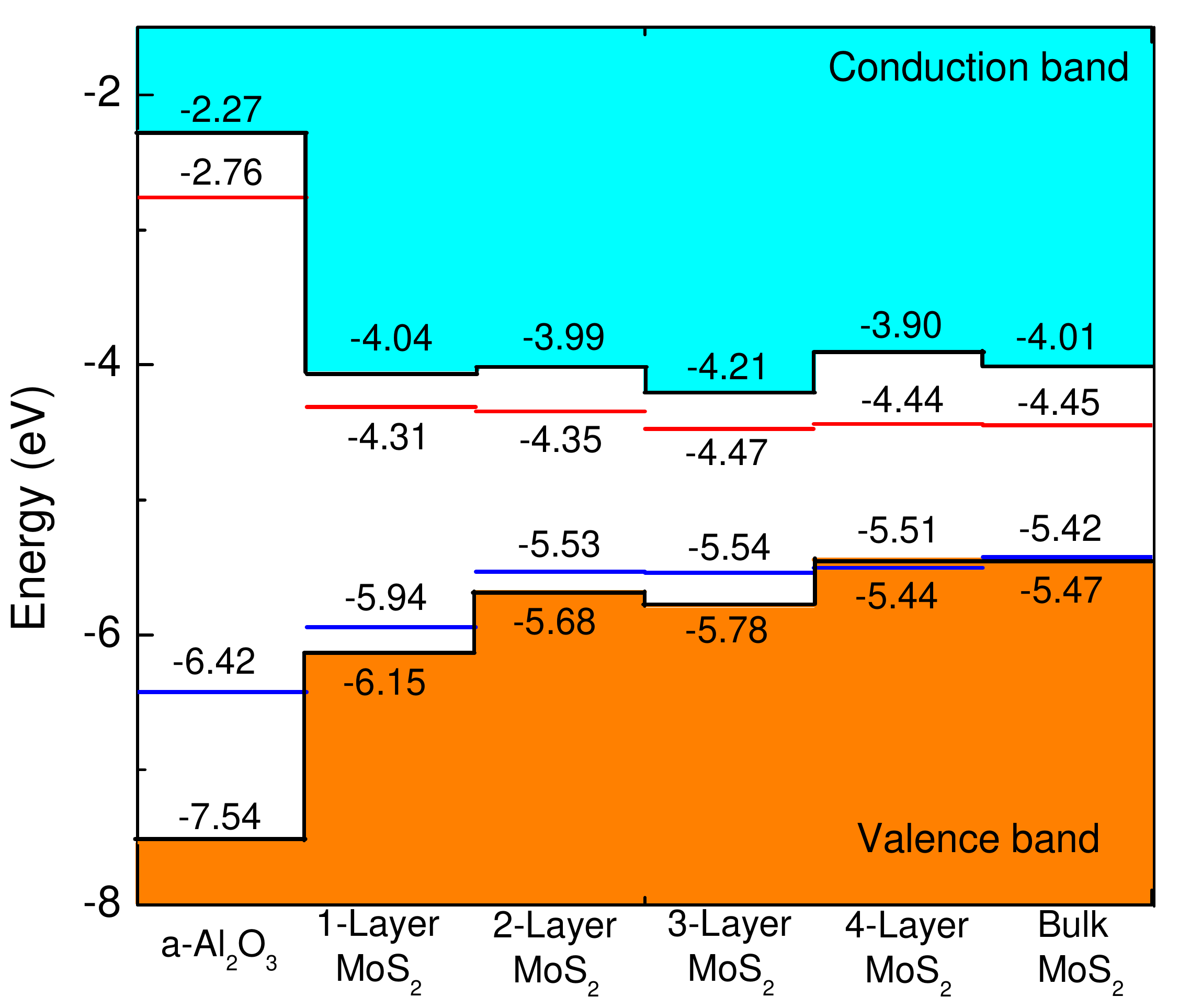}
\caption{Band alignment for $\rm Al_2O_3$ and $\rm MoS_2$ interface. The VBM and CBM positions of a-$\rm Al_2O_3$ and $\rm MoS_2$ are shown in Fig. Solid black lines represent the HSE results; Red and blue lines represent the GGA-PBE results.}
\label{Fig:4}
\end{figure}

\begin{figure}[htp]
\centering
\includegraphics[width=8.5 cm]{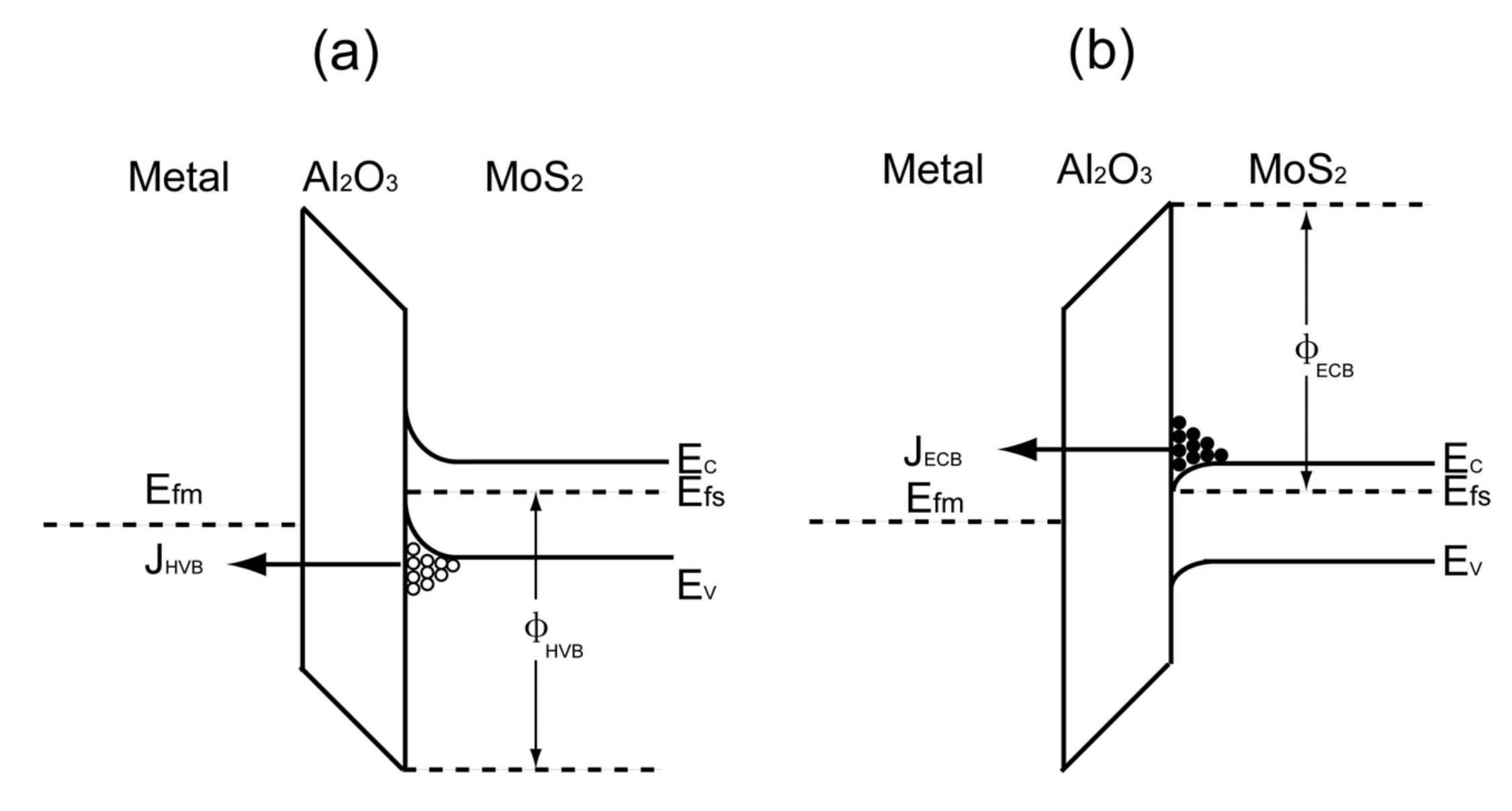}
\caption{Energy band diagrams for metal/$\rm AlO_2$/$\rm MoS_2$ MOS, (a) negative voltage conditions, (b) positive voltage conditions.}
\label{Fig:5}
\end{figure}
\par
Figure 4 shows band alignment between a-$\rm Al_2O_3$ and $\rm MoS_2$. In the following, we discuss the VBO and CBO at a-$\rm Al_2O_3$/$\rm MoS_2$ interface in Fig. 04 based on HSE and GGA-PBE calculations. The VBO and CBO based on HSE calculations are 1.39 eV and 1.77 eV for 1-layer $\rm MoS_2$, 1.86 eV and 1.72 eV for 2-layer $\rm MoS_2$, 1.76 eV and 1.94 eV for 3-layer $\rm MoS_2$, 2.10 eV and 1.63 eV for 4-layer $\rm MoS_2$, and 2.07 eV and 1.74 eV for bulk $\rm MoS_2$. Previous investigations have indicated that VBO and CBO can be affected by the thickness of $\rm MoS_2$.\cite{Nishiguchi2015Observing} The VBO and CBO based on GGA-PBE calculations are 0.48 eV and 1.55 eV for 1-layer $\rm MoS_2$, 0.89 eV and 1.59 eV for 2-layer $\rm MoS_2$, 0.88 eV and 1.71 eV for 3-layer $\rm MoS_2$, 0.91 eV and 1.68 eV for 4-layer $\rm MoS_2$, and 1.00 eV and 1.69 eV for bulk $\rm MoS_2$. GGA-PBE underestimates the band gap, which leads to inaccurate VBO and CBO.\cite{lin2013chemical,Yang2014First} The VBM and CBM positions of a-$\rm Al_2O_3$ and $\rm MoS_2$ are shown in Fig. 04. The VBM position of monolayer $\rm MoS_2$ is -5.94 eV for GGA-PBE, and -6.15 eV for HSE, which is close to previous results of -5.87 eV and -6.27 eV.\cite{Kang2013Band} It is noted that VBM position moves upward as the number of $\rm MoS_2$ layers is increased.
\par
Figure 5 exhibits energy band diagrams for metal/a-$\rm Al_2O_3$/n-type $\rm MoS_2$ MOS under negative and positive voltages. The energy band near $\rm MoS_2$  will bend upwards as the negative voltage is applied between the metal and $\rm MoS_2$.\cite{yeo2000direct} The $\rm MoS_2$ surface layer changes from the majority carrier depletion to the minority carrier inversion with the increase of the voltage. In this case, the hole tunneling barrier of $\phi_{HVB}$ is determined by VBO. Leakage current is expressed as $J_{HVB}$. The energy band near $\rm MoS_2$ will bend down as the positive voltage is applied to devices. The $\rm MoS_2$ surface layer is in majority carrier accumulation region with the increase of the positive voltage. In this case, the electron tunneling barrier of $\phi_{ECB}$ is determined by CBO. The leakage current is written as $J_{ECB}$. Similar analysis is also suitable for metal/a-$\rm Al_2O_3$/p-type $\rm MoS_2$ MOS. In our model, the carrier tunneling current is closely related to the VBO and CBO. The band offset is sensitive to thickness of $\rm MoS_2$, which affects the leakage current of the device to a certain extent. At present, it is generally believed that VBO or CBO in ideal MOS device should be greater than 1 eV.\cite{das2010characterization, hong1999epitaxial,van2013defects,sun2008impact,sun2008effect,sun2007comparative} It is noted that the VBO or CBO calculated by HSE for different $\rm MoS_2$ thickness is greater than 1 eV. Therefore, we believe that metal/a-$\rm Al_2O_3$/$\rm MoS_2$ MOS can be an ideal device.
\subsection{Impact of biaxial strain on the band alignment}
\begin{figure*}
\centering
\includegraphics[width=16 cm]{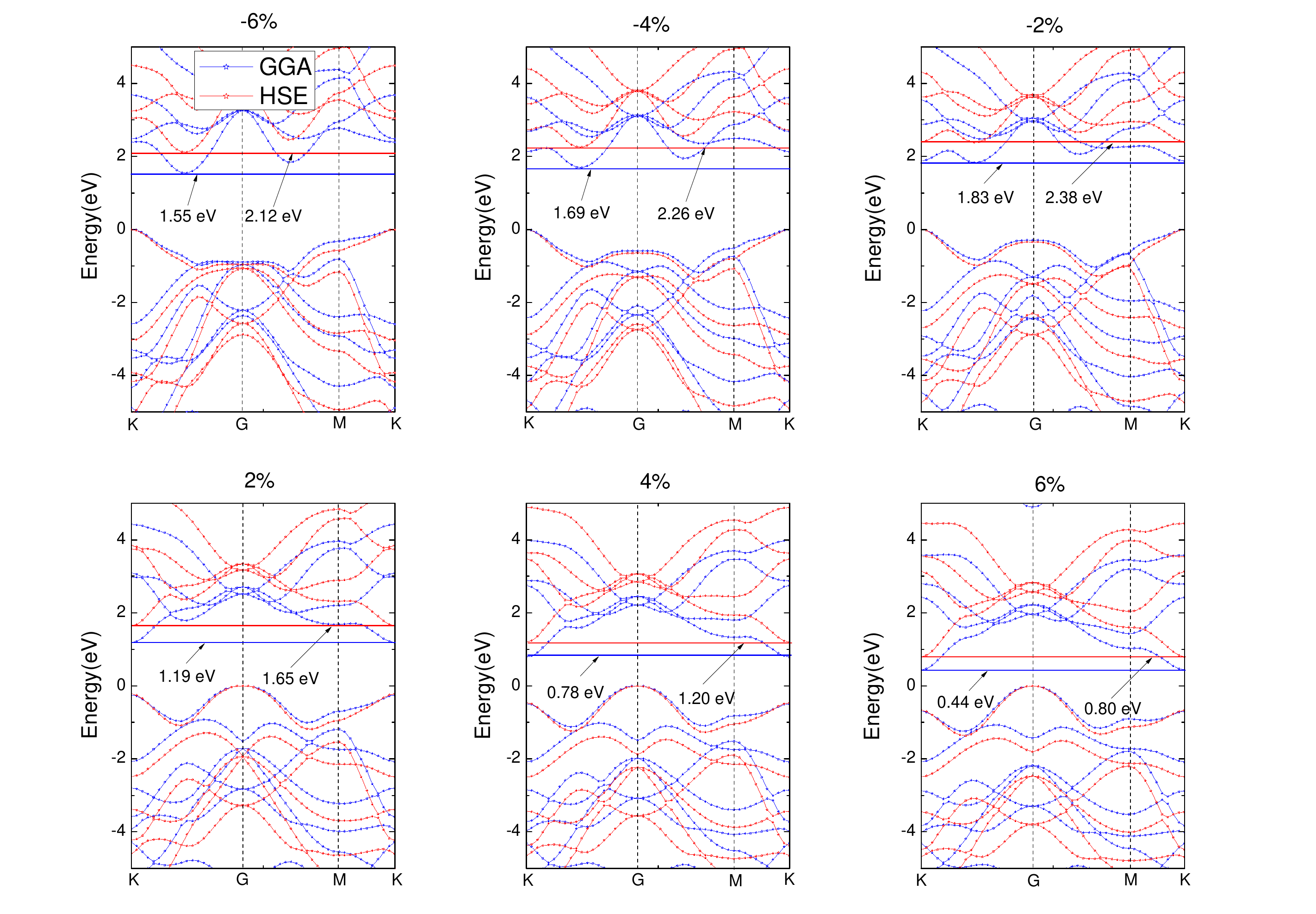}
\caption{Band structures for monolayer $\rm MoS_2$ under biaxial strain in the range from -6\% to 6\%. The bulk and red point connections represent GGA and HSE results. The VBM is set to zero in order to check the band gap. The blue and red horizontal solid line in each panel indicates the CBM calculated by GGA and HSE functional. The arrows indicate the band gap values. Special points in Brillouin zone are set as K (-0.33 0.67 0); G (0 0 0); M (0 0.50 0).}
\label{Fig:6}
\end{figure*}
\begin{figure}
\centering
\includegraphics[width=8 cm]{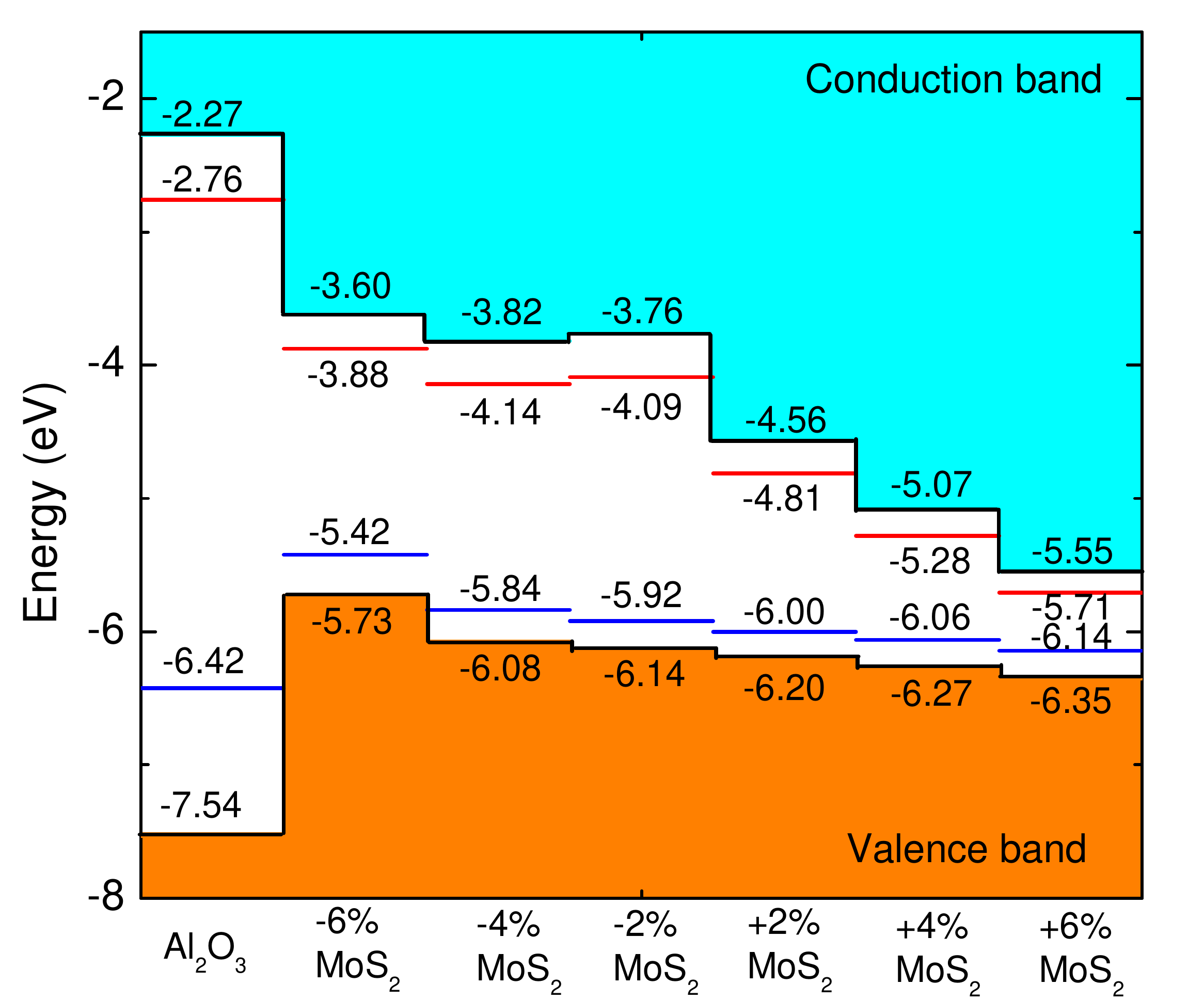}
\caption{Band alignment for a-$\rm Al_2O_3$/monolayer $\rm MoS_2$ interface under biaxial strain in the range from -6\% to 6\%. The VBM and CBM position of a-$\rm Al_2O_3$ and monolayer $\rm MoS_2$ are shown in Fig. Solid black lines represent the HSE results; Red and blue lines represent the GGA-PBE results.}
\label{Fig:7}
\end{figure}
\par
The biaxial strain of monolayer $\rm MoS_2$ is defined as $\varepsilon= \frac{c-c_0}{c_0}\times 100\%$,\cite{kaloni2013hole} where $c$ and $c_0$ are lattice constants of strained and non strained $\rm MoS_2$. $\varepsilon>0$ and $\varepsilon<0$ corresponds to the tensile and compressive strain, respectively. Strain engineering may improve physical performances of semiconductors, in particular, their transport properties.\cite{liu2015strain,yu2015phase,chang2015hole,zhang2009physical,fiori2013strain,cheng2013uniaxial,chern2014record} This inspires us to study the effect of biaxial strain on monolayer $\rm MoS_2$. The lattice strain is set in the range from -6\% to 6\% with the interval of 2\%. Figure 6 presents the band structures for monolayer $\rm MoS_2$ under biaxial strain. The monolayer $\rm MoS_2$ is change to be a indirect band gap semiconductor under biaxial compressive strain, the VBM at K-point, and the CBM between K-point and G-point. The band gaps of GGA and HSE are increased to 1.83 eV and 2.38 eV for -2\%, 1.69 eV and 2.26 eV for -4\%, and 1.55 eV and 2.12 eV for -6\%. The band gap first achieves the maximum value at $\varepsilon$=-2\%, and then gradually decreases with biaxial compressive strain. Similar results are also found on GaAs.\cite{Shi2017First} The band structures of HSE are similar in shape with GGA-PBE, and the band gap values are increased by HSE functional. Interestingly, under biaxial tensile strain, the position of VBM shifts to G-point, and the CBM to K-point. The band gaps are decreased to 1.19 eV and 1.65 eV for 2\%, 0.78 eV and 1.20 eV for 4\%, 0.44 eV and 0.80 eV for 6\%. It is noted that band gaps for monolayer $\rm MoS_2$ will decrease with biaxial tensile strain, which is in agreement with previous investigation. \cite{Nguyen2016Effect,Lanzillo2016Band}
\par
Figure 7 presents band alignment for a-$\rm Al_2O_3$/monolayer $\rm MoS_2$ interface under biaxial strain. The VBM and CBM position of a-$\rm Al_2O_3$ and monolayer $\rm MoS_2$ are shown in Fig. 7. The VBM of monolayer $\rm MoS_2$ consists of $d_{x^2-y^2}$, $d_{xy}$ orbits of Mo and $p_{x}$, $p_{y}$ orbits of S.\cite{Liu2015Electronic} Its position originates from the repulsion between $d_{x^2-y^2}$, $d_{xy}$ orbits of Mo and $p_{x}$, $p_{y}$ orbits of S. The CBM of monolayer $\rm MoS_2$ consists of $d_{z^2}$ orbit of Mo and $p_{x}$, $p_{y}$ orbits of S. Its position originates from the repulsion between $d_{z^2}$ orbit of Mo and $p_{x}$, $p_{y}$ orbits of S. This repulsion is closely related to the overlap of the d orbits of Mo and p orbits of S, and their difference in energy. The overlap is gradually weakened as the biaxial strain changes from -6\% to 6\%. Therefore, the VBM and CBM of monolayer $\rm MoS_2$ move downward as the strain changes from compressive strain to tensile strain. For GGA-PBE calculations, the VBO and CBO are 1.00 and 1.12 eV for $\varepsilon$=-6\%, 0.58 and 1.38 eV for $\varepsilon$=-4\%, 0.50 and 1.33 eV for $\varepsilon$=-2\%, 0.42 and 2.05 eV for $\varepsilon$=2\%, 0.36 and 2.52 eV for $\varepsilon$=4\%, 0.28 and 2.95 eV for $\varepsilon$=6\%. For HSE calculations, the VBO and CBO are increased to 1.81 and 1.33 eV for $\varepsilon$=-6\%, 1.46 and 1.55 eV for $\varepsilon$=-4\%, 1.40 and 1.49 eV for $\varepsilon$=-2\%, 1.34 and 2.29 eV for $\varepsilon$=2\%, 1.27 and 2.80 eV for $\varepsilon$=4\%, 1.19 and 3.28 eV for $\varepsilon$=6\%. It is found from the results that the appropriate biaxial tensile strain in monolayer $\rm MoS_2$ can increase the CBO, while the change of VBO is smaller, which effectively suppresses leakage current of the devices. Previously, Tabatabaei et al. \cite{mohammad2013a} have found that performance of $\rm MoS_2$ field effect transistor can be improved significantly by biaxial strain, which is consistent with the our investigation.

\section{Conclusions}
\par
The generation process of amorphous-$\rm Al_2O_3$ sample is described by molecular dynamics and geometric optimization. The averaged bond lengths of O-O, Al-O and Al-Al for our amorphous-$\rm Al_2O_3$ sample are agreement with previous simulations and experiments. The results show that our sample is close to the actual situation.
\par
The band alignment of oxide and semiconductor can be obtained by building both surface and interface methods. We have verified the results by building $\rm SrTiO_3$/$\rm TiO_2$ interface. It is found that the surface and interface calculations can give similar results.
\par
In order to avoid the waste of calculation time and AEP distortion, we realize band alignment of amorphous-$\rm Al_2O_3$/$\rm MoS_2$ by building both amorphous-$\rm Al_2O_3$ and $\rm MoS_2$ surfaces. For HSE calculations, the VBO and CBO are 1.39 eV and 1.77 eV for 1-layer $\rm MoS_2$, 1.86 eV and 1.72 eV for 2-layer $\rm MoS_2$, 1.76 eV and 1.94 eV for 3-layer $\rm MoS_2$, 2.10 eV and 1.63 eV for 4-layer $\rm MoS_2$, and 2.07 eV and 1.74 eV for bulk $\rm MoS_2$.
The VBO and CBO can be changed by the thickness because the band gap of $\rm MoS_2$ is sensitive to its thickness. The effect of VBO and CBO on the leakage current is also analyzed.
\par
The VBO and CBO based on HSE calculation are increased to 1.81 and 1.33 eV for $\varepsilon$=-6\%, 1.46 and 1.55 eV for $\varepsilon$=-4\%, 1.40 and 1.49 eV for $\varepsilon$=-2\%, 1.34 and 2.29 eV for $\varepsilon$=2\%, 1.27 and 2.80 eV for $\varepsilon$=4\%, 1.19 and 3.28 eV for $\varepsilon$=6\%. The positions of VBM and CBM for monolayer $\rm MoS_2$ move downward as the strain changes from $\varepsilon$=-6\% to $\varepsilon$=6\%. It is noted that the VBO and CBO are larger than 1 eV, indicating that metal/amorphous-$\rm Al_2O_3$/$\rm MoS_2$ is an ideal MOS device.

\begin{acknowledgements}
The work is supported by the National Natural Science Foundation of China under Grant No. 11547182 and No. 11674037.
\end{acknowledgements}


\begin{thebibliography}{85}%
\makeatletter
\providecommand \@ifxundefined [1]{%
 \@ifx{#1\undefined}
}%
\providecommand \@ifnum [1]{%
 \ifnum #1\expandafter \@firstoftwo
 \else \expandafter \@secondoftwo
 \fi
}%
\providecommand \@ifx [1]{%
 \ifx #1\expandafter \@firstoftwo
 \else \expandafter \@secondoftwo
 \fi
}%
\providecommand \natexlab [1]{#1}%
\providecommand \enquote  [1]{``#1''}%
\providecommand \bibnamefont  [1]{#1}%
\providecommand \bibfnamefont [1]{#1}%
\providecommand \citenamefont [1]{#1}%
\providecommand \href@noop [0]{\@secondoftwo}%
\providecommand \href [0]{\begingroup \@sanitize@url \@href}%
\providecommand \@href[1]{\@@startlink{#1}\@@href}%
\providecommand \@@href[1]{\endgroup#1\@@endlink}%
\providecommand \@sanitize@url [0]{\catcode `\\12\catcode `\$12\catcode
  `\&12\catcode `\#12\catcode `\^12\catcode `\_12\catcode `\%12\relax}%
\providecommand \@@startlink[1]{}%
\providecommand \@@endlink[0]{}%
\providecommand \url  [0]{\begingroup\@sanitize@url \@url }%
\providecommand \@url [1]{\endgroup\@href {#1}{\urlprefix }}%
\providecommand \urlprefix  [0]{URL }%
\providecommand \Eprint [0]{\href }%
\providecommand \doibase [0]{http://dx.doi.org/}%
\providecommand \selectlanguage [0]{\@gobble}%
\providecommand \bibinfo  [0]{\@secondoftwo}%
\providecommand \bibfield  [0]{\@secondoftwo}%
\providecommand \translation [1]{[#1]}%
\providecommand \BibitemOpen [0]{}%
\providecommand \bibitemStop [0]{}%
\providecommand \bibitemNoStop [0]{.\EOS\space}%
\providecommand \EOS [0]{\spacefactor3000\relax}%
\providecommand \BibitemShut  [1]{\csname bibitem#1\endcsname}%
\let\auto@bib@innerbib\@empty
\bibitem [{\citenamefont {Moore}(1998)}]{moore1998cramming}%
  \BibitemOpen
  \bibfield  {author} {\bibinfo {author} {\bibfnamefont {G.~E.}\ \bibnamefont
  {Moore}},\ }\href@noop {} {\bibfield  {journal} {\bibinfo  {journal}
  {Proceedings of the IEEE}\ }\textbf {\bibinfo {volume} {86}},\ \bibinfo
  {pages} {82} (\bibinfo {year} {1998})}\BibitemShut {NoStop}%
\bibitem [{\citenamefont {He}\ \emph {et~al.}(2013)\citenamefont {He},
  \citenamefont {Chen},\ and\ \citenamefont {Sun}}]{he2013interface}%
  \BibitemOpen
  \bibfield  {author} {\bibinfo {author} {\bibfnamefont {G.}~\bibnamefont
  {He}}, \bibinfo {author} {\bibfnamefont {X.}~\bibnamefont {Chen}}, \ and\
  \bibinfo {author} {\bibfnamefont {Z.}~\bibnamefont {Sun}},\ }\href@noop {}
  {\bibfield  {journal} {\bibinfo  {journal} {Surface Science Reports}\
  }\textbf {\bibinfo {volume} {68}},\ \bibinfo {pages} {68} (\bibinfo {year}
  {2013})}\BibitemShut {NoStop}%
\bibitem [{\citenamefont {He}\ \emph {et~al.}(2011)\citenamefont {He},
  \citenamefont {Zhu}, \citenamefont {Sun}, \citenamefont {Wan},\ and\
  \citenamefont {Zhang}}]{he2011integrations}%
  \BibitemOpen
  \bibfield  {author} {\bibinfo {author} {\bibfnamefont {G.}~\bibnamefont
  {He}}, \bibinfo {author} {\bibfnamefont {L.}~\bibnamefont {Zhu}}, \bibinfo
  {author} {\bibfnamefont {Z.}~\bibnamefont {Sun}}, \bibinfo {author}
  {\bibfnamefont {Q.}~\bibnamefont {Wan}}, \ and\ \bibinfo {author}
  {\bibfnamefont {L.}~\bibnamefont {Zhang}},\ }\href@noop {} {\bibfield
  {journal} {\bibinfo  {journal} {Progress in Materials Science}\ }\textbf
  {\bibinfo {volume} {56}},\ \bibinfo {pages} {475} (\bibinfo {year}
  {2011})}\BibitemShut {NoStop}%
\bibitem [{\citenamefont {Fang}\ \emph {et~al.}(2012)\citenamefont {Fang},
  \citenamefont {Hodson}, \citenamefont {Liu}, \citenamefont {Fang},
  \citenamefont {Potter},\ and\ \citenamefont {Gunn}}]{fang2012preliminary}%
  \BibitemOpen
  \bibfield  {author} {\bibinfo {author} {\bibfnamefont {Q.~F.}\ \bibnamefont
  {Fang}}, \bibinfo {author} {\bibfnamefont {C.}~\bibnamefont {Hodson}},
  \bibinfo {author} {\bibfnamefont {M.}~\bibnamefont {Liu}}, \bibinfo {author}
  {\bibfnamefont {Z.~W.}\ \bibnamefont {Fang}}, \bibinfo {author}
  {\bibfnamefont {R.}~\bibnamefont {Potter}}, \ and\ \bibinfo {author}
  {\bibfnamefont {R.}~\bibnamefont {Gunn}},\ }\href@noop {} {\bibfield
  {journal} {\bibinfo  {journal} {Physics Procedia}\ }\textbf {\bibinfo
  {volume} {32}},\ \bibinfo {pages} {379} (\bibinfo {year} {2012})}\BibitemShut
  {NoStop}%
\bibitem [{\citenamefont {Park}\ \emph {et~al.}(1996)\citenamefont {Park},
  \citenamefont {Diatezua}, \citenamefont {Chen}, \citenamefont {Mohammad},\
  and\ \citenamefont {Morko{\c{c}}}}]{park1996characteristics}%
  \BibitemOpen
  \bibfield  {author} {\bibinfo {author} {\bibfnamefont {D.-G.}\ \bibnamefont
  {Park}}, \bibinfo {author} {\bibfnamefont {D.~M.}\ \bibnamefont {Diatezua}},
  \bibinfo {author} {\bibfnamefont {Z.}~\bibnamefont {Chen}}, \bibinfo {author}
  {\bibfnamefont {S.~N.}\ \bibnamefont {Mohammad}}, \ and\ \bibinfo {author}
  {\bibfnamefont {H.}~\bibnamefont {Morko{\c{c}}}},\ }\href@noop {} {\bibfield
  {journal} {\bibinfo  {journal} {Applied physics letters}\ }\textbf {\bibinfo
  {volume} {69}},\ \bibinfo {pages} {3025} (\bibinfo {year}
  {1996})}\BibitemShut {NoStop}%
\bibitem [{\citenamefont {Hong}\ \emph {et~al.}(1999)\citenamefont {Hong},
  \citenamefont {Kwo}, \citenamefont {Kortan}, \citenamefont {Mannaerts},\ and\
  \citenamefont {Sergent}}]{hong1999epitaxial}%
  \BibitemOpen
  \bibfield  {author} {\bibinfo {author} {\bibfnamefont {M.}~\bibnamefont
  {Hong}}, \bibinfo {author} {\bibfnamefont {J.}~\bibnamefont {Kwo}}, \bibinfo
  {author} {\bibfnamefont {A.}~\bibnamefont {Kortan}}, \bibinfo {author}
  {\bibfnamefont {J.}~\bibnamefont {Mannaerts}}, \ and\ \bibinfo {author}
  {\bibfnamefont {A.}~\bibnamefont {Sergent}},\ }\href@noop {} {\bibfield
  {journal} {\bibinfo  {journal} {Science}\ }\textbf {\bibinfo {volume}
  {283}},\ \bibinfo {pages} {1897} (\bibinfo {year} {1999})}\BibitemShut
  {NoStop}%
\bibitem [{\citenamefont {Chiu}\ \emph {et~al.}(2005)\citenamefont {Chiu},
  \citenamefont {Chou},\ and\ \citenamefont {Lee}}]{chiu2005electrical}%
  \BibitemOpen
  \bibfield  {author} {\bibinfo {author} {\bibfnamefont {F.~C.}\ \bibnamefont
  {Chiu}}, \bibinfo {author} {\bibfnamefont {H.~W.}\ \bibnamefont {Chou}}, \
  and\ \bibinfo {author} {\bibfnamefont {J.~Y.}\ \bibnamefont {Lee}},\
  }\href@noop {} {\bibfield  {journal} {\bibinfo  {journal} {Journal of Applied
  Physics}\ }\textbf {\bibinfo {volume} {97}},\ \bibinfo {pages} {103503 }
  (\bibinfo {year} {2005})}\BibitemShut {NoStop}%
\bibitem [{\citenamefont {Wu}\ \emph {et~al.}(2015)\citenamefont {Wu},
  \citenamefont {Chen}, \citenamefont {Lin}, \citenamefont {Cheng},
  \citenamefont {Hsu}, \citenamefont {Kwo},\ and\ \citenamefont
  {Hong}}]{wu2015single}%
  \BibitemOpen
  \bibfield  {author} {\bibinfo {author} {\bibfnamefont {S.}~\bibnamefont
  {Wu}}, \bibinfo {author} {\bibfnamefont {K.}~\bibnamefont {Chen}}, \bibinfo
  {author} {\bibfnamefont {Y.}~\bibnamefont {Lin}}, \bibinfo {author}
  {\bibfnamefont {C.}~\bibnamefont {Cheng}}, \bibinfo {author} {\bibfnamefont
  {C.}~\bibnamefont {Hsu}}, \bibinfo {author} {\bibfnamefont {J.}~\bibnamefont
  {Kwo}}, \ and\ \bibinfo {author} {\bibfnamefont {M.}~\bibnamefont {Hong}},\
  }\href@noop {} {\bibfield  {journal} {\bibinfo  {journal} {Microelectronic
  Engineering}\ }\textbf {\bibinfo {volume} {147}},\ \bibinfo {pages} {310}
  (\bibinfo {year} {2015})}\BibitemShut {NoStop}%
\bibitem [{\citenamefont {Cai}\ \emph {et~al.}(2007)\citenamefont {Cai},
  \citenamefont {Stone}, \citenamefont {Pelz}, \citenamefont {Edge},\ and\
  \citenamefont {Schlom}}]{cai2007conduction}%
  \BibitemOpen
  \bibfield  {author} {\bibinfo {author} {\bibfnamefont {W.}~\bibnamefont
  {Cai}}, \bibinfo {author} {\bibfnamefont {S.}~\bibnamefont {Stone}}, \bibinfo
  {author} {\bibfnamefont {J.~P.}\ \bibnamefont {Pelz}}, \bibinfo {author}
  {\bibfnamefont {L.~F.}\ \bibnamefont {Edge}}, \ and\ \bibinfo {author}
  {\bibfnamefont {D.~G.}\ \bibnamefont {Schlom}},\ }\href@noop {} {\bibfield
  {journal} {\bibinfo  {journal} {Applied Physics Letters}\ }\textbf {\bibinfo
  {volume} {91}},\ \bibinfo {pages} {042901} (\bibinfo {year}
  {2007})}\BibitemShut {NoStop}%
\bibitem [{\citenamefont {Hsieh}\ \emph {et~al.}(2008)\citenamefont {Hsieh},
  \citenamefont {Chang}, \citenamefont {Chien}, \citenamefont {Chou},
  \citenamefont {Chen},\ and\ \citenamefont {Chen}}]{hsieh2008coaxial}%
  \BibitemOpen
  \bibfield  {author} {\bibinfo {author} {\bibfnamefont {C.}~\bibnamefont
  {Hsieh}}, \bibinfo {author} {\bibfnamefont {M.}~\bibnamefont {Chang}},
  \bibinfo {author} {\bibfnamefont {Y.}~\bibnamefont {Chien}}, \bibinfo
  {author} {\bibfnamefont {L.}~\bibnamefont {Chou}}, \bibinfo {author}
  {\bibfnamefont {L.}~\bibnamefont {Chen}}, \ and\ \bibinfo {author}
  {\bibfnamefont {C.}~\bibnamefont {Chen}},\ }\href@noop {} {\bibfield
  {journal} {\bibinfo  {journal} {Nano Letters}\ }\textbf {\bibinfo {volume}
  {8}},\ \bibinfo {pages} {3288} (\bibinfo {year} {2008})}\BibitemShut
  {NoStop}%
\bibitem [{\citenamefont {Xiong}\ and\ \citenamefont
  {Robertson}(2009)}]{xiong2009electronic}%
  \BibitemOpen
  \bibfield  {author} {\bibinfo {author} {\bibfnamefont {K.}~\bibnamefont
  {Xiong}}\ and\ \bibinfo {author} {\bibfnamefont {J.}~\bibnamefont
  {Robertson}},\ }\href@noop {} {\bibfield  {journal} {\bibinfo  {journal}
  {Microelectronic Engineering}\ }\textbf {\bibinfo {volume} {86}},\ \bibinfo
  {pages} {1672} (\bibinfo {year} {2009})}\BibitemShut {NoStop}%
\bibitem [{\citenamefont {Guo}\ \emph {et~al.}(2013)\citenamefont {Guo},
  \citenamefont {Lin},\ and\ \citenamefont {Robertson}}]{guo2013nitrogen}%
  \BibitemOpen
  \bibfield  {author} {\bibinfo {author} {\bibfnamefont {Y.}~\bibnamefont
  {Guo}}, \bibinfo {author} {\bibfnamefont {L.}~\bibnamefont {Lin}}, \ and\
  \bibinfo {author} {\bibfnamefont {J.}~\bibnamefont {Robertson}},\ }\href@noop
  {} {\bibfield  {journal} {\bibinfo  {journal} {Applied Physics Letters}\
  }\textbf {\bibinfo {volume} {102}},\ \bibinfo {pages} {091606} (\bibinfo
  {year} {2013})}\BibitemShut {NoStop}%
\bibitem [{\citenamefont {Choi}\ \emph
  {et~al.}(2013{\natexlab{a}})\citenamefont {Choi}, \citenamefont {Lyons},
  \citenamefont {Janotti},\ and\ \citenamefont {De~Walle}}]{choi2013impact}%
  \BibitemOpen
  \bibfield  {author} {\bibinfo {author} {\bibfnamefont {M.}~\bibnamefont
  {Choi}}, \bibinfo {author} {\bibfnamefont {J.~L.}\ \bibnamefont {Lyons}},
  \bibinfo {author} {\bibfnamefont {A.}~\bibnamefont {Janotti}}, \ and\
  \bibinfo {author} {\bibfnamefont {C.~G.~V.}\ \bibnamefont {De~Walle}},\
  }\href@noop {} {\bibfield  {journal} {\bibinfo  {journal} {Applied Physics
  Letters}\ }\textbf {\bibinfo {volume} {102}},\ \bibinfo {pages} {142902}
  (\bibinfo {year} {2013}{\natexlab{a}})}\BibitemShut {NoStop}%
\bibitem [{\citenamefont {Hoex}\ \emph {et~al.}(2008)\citenamefont {Hoex},
  \citenamefont {Gielis}, \citenamefont {De~Sanden},\ and\ \citenamefont
  {Kessels}}]{hoex2008on}%
  \BibitemOpen
  \bibfield  {author} {\bibinfo {author} {\bibfnamefont {B.}~\bibnamefont
  {Hoex}}, \bibinfo {author} {\bibfnamefont {J.~J.~H.}\ \bibnamefont {Gielis}},
  \bibinfo {author} {\bibfnamefont {M.~C. M.~V.}\ \bibnamefont {De~Sanden}}, \
  and\ \bibinfo {author} {\bibfnamefont {W.~M.~M.}\ \bibnamefont {Kessels}},\
  }\href@noop {} {\bibfield  {journal} {\bibinfo  {journal} {Journal of Applied
  Physics}\ }\textbf {\bibinfo {volume} {104}},\ \bibinfo {pages} {113703}
  (\bibinfo {year} {2008})}\BibitemShut {NoStop}%
\bibitem [{\citenamefont {Werner}\ \emph {et~al.}(2011)\citenamefont {Werner},
  \citenamefont {Veith}, \citenamefont {Zielke}, \citenamefont {Kuhnemund},
  \citenamefont {Tegenkamp}, \citenamefont {Seibt}, \citenamefont {Brendel},\
  and\ \citenamefont {Schmidt}}]{werner2011electronic}%
  \BibitemOpen
  \bibfield  {author} {\bibinfo {author} {\bibfnamefont {F.}~\bibnamefont
  {Werner}}, \bibinfo {author} {\bibfnamefont {B.}~\bibnamefont {Veith}},
  \bibinfo {author} {\bibfnamefont {D.}~\bibnamefont {Zielke}}, \bibinfo
  {author} {\bibfnamefont {L.}~\bibnamefont {Kuhnemund}}, \bibinfo {author}
  {\bibfnamefont {C.}~\bibnamefont {Tegenkamp}}, \bibinfo {author}
  {\bibfnamefont {M.}~\bibnamefont {Seibt}}, \bibinfo {author} {\bibfnamefont
  {R.}~\bibnamefont {Brendel}}, \ and\ \bibinfo {author} {\bibfnamefont
  {J.}~\bibnamefont {Schmidt}},\ }\href@noop {} {\bibfield  {journal} {\bibinfo
   {journal} {Journal of Applied Physics}\ }\textbf {\bibinfo {volume} {109}},\
  \bibinfo {pages} {113701} (\bibinfo {year} {2011})}\BibitemShut {NoStop}%
\bibitem [{\citenamefont {Lin}\ \emph {et~al.}(2013)\citenamefont {Lin},
  \citenamefont {Guo}, \citenamefont {Gillen},\ and\ \citenamefont
  {Robertson}}]{lin2013chemical}%
  \BibitemOpen
  \bibfield  {author} {\bibinfo {author} {\bibfnamefont {L.}~\bibnamefont
  {Lin}}, \bibinfo {author} {\bibfnamefont {Y.}~\bibnamefont {Guo}}, \bibinfo
  {author} {\bibfnamefont {R.}~\bibnamefont {Gillen}}, \ and\ \bibinfo {author}
  {\bibfnamefont {J.}~\bibnamefont {Robertson}},\ }\href@noop {} {\bibfield
  {journal} {\bibinfo  {journal} {Journal of Applied Physics}\ }\textbf
  {\bibinfo {volume} {113}},\ \bibinfo {pages} {134103} (\bibinfo {year}
  {2013})}\BibitemShut {NoStop}%
\bibitem [{\citenamefont {Suh}\ \emph {et~al.}(2013)\citenamefont {Suh},
  \citenamefont {Choi},\ and\ \citenamefont {Weber}}]{suh2013al2o3/tio2}%
  \BibitemOpen
  \bibfield  {author} {\bibinfo {author} {\bibfnamefont {D.}~\bibnamefont
  {Suh}}, \bibinfo {author} {\bibfnamefont {D.}~\bibnamefont {Choi}}, \ and\
  \bibinfo {author} {\bibfnamefont {K.}~\bibnamefont {Weber}},\ }\href@noop {}
  {\bibfield  {journal} {\bibinfo  {journal} {Journal of Applied Physics}\
  }\textbf {\bibinfo {volume} {114}},\ \bibinfo {pages} {154107} (\bibinfo
  {year} {2013})}\BibitemShut {NoStop}%
\bibitem [{\citenamefont {Liu}\ \emph {et~al.}(2013{\natexlab{a}})\citenamefont
  {Liu}, \citenamefont {Liao}, \citenamefont {Imura}, \citenamefont {Oosato},
  \citenamefont {Watanabe}, \citenamefont {Tanaka}, \citenamefont {Iwai},
  \citenamefont {Koide},\ and\ \citenamefont {Ibaraki}}]{liu2013interfacial}%
  \BibitemOpen
  \bibfield  {author} {\bibinfo {author} {\bibfnamefont {J.}~\bibnamefont
  {Liu}}, \bibinfo {author} {\bibfnamefont {M.~Y.}\ \bibnamefont {Liao}},
  \bibinfo {author} {\bibfnamefont {M.}~\bibnamefont {Imura}}, \bibinfo
  {author} {\bibfnamefont {H.}~\bibnamefont {Oosato}}, \bibinfo {author}
  {\bibfnamefont {E.}~\bibnamefont {Watanabe}}, \bibinfo {author}
  {\bibfnamefont {A.}~\bibnamefont {Tanaka}}, \bibinfo {author} {\bibfnamefont
  {H.}~\bibnamefont {Iwai}}, \bibinfo {author} {\bibfnamefont {Y.}~\bibnamefont
  {Koide}}, \ and\ \bibinfo {author} {\bibfnamefont {N.~P. N. S.~T.}\
  \bibnamefont {Ibaraki}},\ }\href@noop {} {\bibfield  {journal} {\bibinfo
  {journal} {Journal of Applied Physics}\ }\textbf {\bibinfo {volume} {114}},\
  \bibinfo {pages} {084108} (\bibinfo {year} {2013}{\natexlab{a}})}\BibitemShut
  {NoStop}%
\bibitem [{\citenamefont {Choi}\ \emph {et~al.}(2007)\citenamefont {Choi},
  \citenamefont {Cartier}, \citenamefont {Wang}, \citenamefont {Narayanan},\
  and\ \citenamefont {Khare}}]{choi2007dual}%
  \BibitemOpen
  \bibfield  {author} {\bibinfo {author} {\bibfnamefont {C.}~\bibnamefont
  {Choi}}, \bibinfo {author} {\bibfnamefont {E.}~\bibnamefont {Cartier}},
  \bibinfo {author} {\bibfnamefont {Y.}~\bibnamefont {Wang}}, \bibinfo {author}
  {\bibfnamefont {V.}~\bibnamefont {Narayanan}}, \ and\ \bibinfo {author}
  {\bibfnamefont {M.}~\bibnamefont {Khare}},\ }\href@noop {} {\bibfield
  {journal} {\bibinfo  {journal} {Microelectronic Engineering}\ }\textbf
  {\bibinfo {volume} {84}},\ \bibinfo {pages} {2217} (\bibinfo {year}
  {2007})}\BibitemShut {NoStop}%
\bibitem [{\citenamefont {Olyaei}\ \emph {et~al.}(2012)\citenamefont {Olyaei},
  \citenamefont {Malm}, \citenamefont {Hellstrom},\ and\ \citenamefont
  {Ostling}}]{olyaei2012low-frequency}%
  \BibitemOpen
  \bibfield  {author} {\bibinfo {author} {\bibfnamefont {M.}~\bibnamefont
  {Olyaei}}, \bibinfo {author} {\bibfnamefont {G.}~\bibnamefont {Malm}},
  \bibinfo {author} {\bibfnamefont {P.}~\bibnamefont {Hellstrom}}, \ and\
  \bibinfo {author} {\bibfnamefont {M.}~\bibnamefont {Ostling}},\ }\href@noop
  {} {\bibfield  {journal} {\bibinfo  {journal} {Solid-state Electronics}\
  }\textbf {\bibinfo {volume} {78}},\ \bibinfo {pages} {51} (\bibinfo {year}
  {2012})}\BibitemShut {NoStop}%
\bibitem [{\citenamefont {Zheng}\ \emph {et~al.}(2007)\citenamefont {Zheng},
  \citenamefont {Ceder}, \citenamefont {Maxisch}, \citenamefont {Chim},\ and\
  \citenamefont {Choi}}]{Zheng2007First}%
  \BibitemOpen
  \bibfield  {author} {\bibinfo {author} {\bibfnamefont {J.~X.}\ \bibnamefont
  {Zheng}}, \bibinfo {author} {\bibfnamefont {G.}~\bibnamefont {Ceder}},
  \bibinfo {author} {\bibfnamefont {T.}~\bibnamefont {Maxisch}}, \bibinfo
  {author} {\bibfnamefont {W.~K.}\ \bibnamefont {Chim}}, \ and\ \bibinfo
  {author} {\bibfnamefont {W.~K.}\ \bibnamefont {Choi}},\ }\href@noop {}
  {\bibfield  {journal} {\bibinfo  {journal} {Physical Review B}\ }\textbf
  {\bibinfo {volume} {75}},\ \bibinfo {pages} {104112} (\bibinfo {year}
  {2007})}\BibitemShut {NoStop}%
\bibitem [{\citenamefont {Kang}\ \emph {et~al.}(2003)\citenamefont {Kang},
  \citenamefont {Lee},\ and\ \citenamefont {Chang}}]{kang2003first-principles}%
  \BibitemOpen
  \bibfield  {author} {\bibinfo {author} {\bibfnamefont {J.}~\bibnamefont
  {Kang}}, \bibinfo {author} {\bibfnamefont {E.}~\bibnamefont {Lee}}, \ and\
  \bibinfo {author} {\bibfnamefont {K.~J.}\ \bibnamefont {Chang}},\ }\href@noop
  {} {\bibfield  {journal} {\bibinfo  {journal} {Physical Review B}\ }\textbf
  {\bibinfo {volume} {68}},\ \bibinfo {pages} {054106} (\bibinfo {year}
  {2003})}\BibitemShut {NoStop}%
\bibitem [{\citenamefont {Xiong}\ \emph {et~al.}(2007)\citenamefont {Xiong},
  \citenamefont {Du}, \citenamefont {Tse},\ and\ \citenamefont
  {Robertson}}]{xiong2007defect}%
  \BibitemOpen
  \bibfield  {author} {\bibinfo {author} {\bibfnamefont {K.}~\bibnamefont
  {Xiong}}, \bibinfo {author} {\bibfnamefont {Y.}~\bibnamefont {Du}}, \bibinfo
  {author} {\bibfnamefont {K.~Y.}\ \bibnamefont {Tse}}, \ and\ \bibinfo
  {author} {\bibfnamefont {J.}~\bibnamefont {Robertson}},\ }\href@noop {}
  {\bibfield  {journal} {\bibinfo  {journal} {Journal of Applied Physics}\
  }\textbf {\bibinfo {volume} {101}},\ \bibinfo {pages} {024101} (\bibinfo
  {year} {2007})}\BibitemShut {NoStop}%
\bibitem [{\citenamefont {Cota}\ \emph {et~al.}(2013)\citenamefont {Cota},
  \citenamefont {Burton}, \citenamefont {Cha¨ªn}, \citenamefont {Pav¨®n},\ and\
  \citenamefont {Alba}}]{cota2013solution}%
  \BibitemOpen
  \bibfield  {author} {\bibinfo {author} {\bibfnamefont {A.}~\bibnamefont
  {Cota}}, \bibinfo {author} {\bibfnamefont {B.~P.}\ \bibnamefont {Burton}},
  \bibinfo {author} {\bibfnamefont {P.}~\bibnamefont {Cha¨ªn}}, \bibinfo
  {author} {\bibfnamefont {E.}~\bibnamefont {Pav¨®n}}, \ and\ \bibinfo {author}
  {\bibfnamefont {M.~D.}\ \bibnamefont {Alba}},\ }\href@noop {} {\bibfield
  {journal} {\bibinfo  {journal} {Journal of Physical Chemistry C}\ }\textbf
  {\bibinfo {volume} {117}},\ \bibinfo {pages} {10013} (\bibinfo {year}
  {2013})}\BibitemShut {NoStop}%
\bibitem [{\citenamefont {Costina}\ and\ \citenamefont
  {Franchy}(2001)}]{Costina2001Band}%
  \BibitemOpen
  \bibfield  {author} {\bibinfo {author} {\bibfnamefont {I.}~\bibnamefont
  {Costina}}\ and\ \bibinfo {author} {\bibfnamefont {R.}~\bibnamefont
  {Franchy}},\ }\href@noop {} {\bibfield  {journal} {\bibinfo  {journal}
  {Applied Physics Letters}\ }\textbf {\bibinfo {volume} {78}},\ \bibinfo
  {pages} {4139} (\bibinfo {year} {2001})}\BibitemShut {NoStop}%
\bibitem [{\citenamefont {Liu}\ \emph {et~al.}(2013{\natexlab{b}})\citenamefont
  {Liu}, \citenamefont {Kobayashi}, \citenamefont {Ohta}, \citenamefont
  {Fujioka},\ and\ \citenamefont {Oshima}}]{Liu2013Electrical}%
  \BibitemOpen
  \bibfield  {author} {\bibinfo {author} {\bibfnamefont {J.~W.}\ \bibnamefont
  {Liu}}, \bibinfo {author} {\bibfnamefont {A.}~\bibnamefont {Kobayashi}},
  \bibinfo {author} {\bibfnamefont {J.}~\bibnamefont {Ohta}}, \bibinfo {author}
  {\bibfnamefont {H.}~\bibnamefont {Fujioka}}, \ and\ \bibinfo {author}
  {\bibfnamefont {M.}~\bibnamefont {Oshima}},\ }\href@noop {} {\bibfield
  {journal} {\bibinfo  {journal} {Applied Physics Letters}\ }\textbf {\bibinfo
  {volume} {103}},\ \bibinfo {pages} {172101} (\bibinfo {year}
  {2013}{\natexlab{b}})}\BibitemShut {NoStop}%
\bibitem [{\citenamefont {Hu}\ \emph {et~al.}(2014)\citenamefont {Hu},
  \citenamefont {Yao}, \citenamefont {Xiao}, \citenamefont {Chen},\ and\
  \citenamefont {Yao}}]{Hu2014Optical}%
  \BibitemOpen
  \bibfield  {author} {\bibinfo {author} {\bibfnamefont {B.}~\bibnamefont
  {Hu}}, \bibinfo {author} {\bibfnamefont {M.}~\bibnamefont {Yao}}, \bibinfo
  {author} {\bibfnamefont {R.}~\bibnamefont {Xiao}}, \bibinfo {author}
  {\bibfnamefont {J.}~\bibnamefont {Chen}}, \ and\ \bibinfo {author}
  {\bibfnamefont {X.}~\bibnamefont {Yao}},\ }\href@noop {} {\bibfield
  {journal} {\bibinfo  {journal} {Ceramics International}\ }\textbf {\bibinfo
  {volume} {40}},\ \bibinfo {pages} {14133} (\bibinfo {year}
  {2014})}\BibitemShut {NoStop}%
\bibitem [{\citenamefont {Chagarov}\ and\ \citenamefont
  {Kummel}(2008)}]{Chagarov2008Generation}%
  \BibitemOpen
  \bibfield  {author} {\bibinfo {author} {\bibfnamefont {E.~A.}\ \bibnamefont
  {Chagarov}}\ and\ \bibinfo {author} {\bibfnamefont {A.~C.}\ \bibnamefont
  {Kummel}},\ }\href@noop {} {\bibfield  {journal} {\bibinfo  {journal} {Ecs
  Transactions}\ }\textbf {\bibinfo {volume} {16}},\ \bibinfo {pages} {8}
  (\bibinfo {year} {2008})}\BibitemShut {NoStop}%
\bibitem [{\citenamefont {Gutierrez}\ and\ \citenamefont
  {Johansson}(2002)}]{Guti2002Molecular}%
  \BibitemOpen
  \bibfield  {author} {\bibinfo {author} {\bibfnamefont {G.}~\bibnamefont
  {Gutierrez}}\ and\ \bibinfo {author} {\bibfnamefont {B.}~\bibnamefont
  {Johansson}},\ }\href@noop {} {\bibfield  {journal} {\bibinfo  {journal}
  {Physical Review B}\ }\textbf {\bibinfo {volume} {65}},\ \bibinfo {pages}
  {104202} (\bibinfo {year} {2002})}\BibitemShut {NoStop}%
\bibitem [{\citenamefont {Lamparter}\ and\ \citenamefont
  {Kniep}(1997)}]{lamparter1997structure}%
  \BibitemOpen
  \bibfield  {author} {\bibinfo {author} {\bibfnamefont {P.}~\bibnamefont
  {Lamparter}}\ and\ \bibinfo {author} {\bibfnamefont {R.}~\bibnamefont
  {Kniep}},\ }\href@noop {} {\bibfield  {journal} {\bibinfo  {journal} {Physica
  B: Condensed Matter}\ }\textbf {\bibinfo {volume} {234}},\ \bibinfo {pages}
  {405} (\bibinfo {year} {1997})}\BibitemShut {NoStop}%
\bibitem [{\citenamefont {Novoselov}\ \emph {et~al.}(2004)\citenamefont
  {Novoselov}, \citenamefont {Geim}, \citenamefont {Morozov}, \citenamefont
  {Jiang}, \citenamefont {Zhang}, \citenamefont {Dubonos}, \citenamefont
  {Grigorieva},\ and\ \citenamefont {Firsov}}]{Novoselov2004Electric}%
  \BibitemOpen
  \bibfield  {author} {\bibinfo {author} {\bibfnamefont {K.~S.}\ \bibnamefont
  {Novoselov}}, \bibinfo {author} {\bibfnamefont {A.~K.}\ \bibnamefont {Geim}},
  \bibinfo {author} {\bibfnamefont {S.~V.}\ \bibnamefont {Morozov}}, \bibinfo
  {author} {\bibfnamefont {D.}~\bibnamefont {Jiang}}, \bibinfo {author}
  {\bibfnamefont {Y.}~\bibnamefont {Zhang}}, \bibinfo {author} {\bibfnamefont
  {S.~V.}\ \bibnamefont {Dubonos}}, \bibinfo {author} {\bibfnamefont {I.~V.}\
  \bibnamefont {Grigorieva}}, \ and\ \bibinfo {author} {\bibfnamefont {A.~A.}\
  \bibnamefont {Firsov}},\ }\href@noop {} {\bibfield  {journal} {\bibinfo
  {journal} {Science}\ }\textbf {\bibinfo {volume} {306}},\ \bibinfo {pages}
  {666} (\bibinfo {year} {2004})}\BibitemShut {NoStop}%
\bibitem [{\citenamefont {Nguyen}\ and\ \citenamefont
  {Hieu}(2016)}]{Nguyen2016Effect}%
  \BibitemOpen
  \bibfield  {author} {\bibinfo {author} {\bibfnamefont {C.~V.}\ \bibnamefont
  {Nguyen}}\ and\ \bibinfo {author} {\bibfnamefont {N.~N.}\ \bibnamefont
  {Hieu}},\ }\href@noop {} {\bibfield  {journal} {\bibinfo  {journal} {Chemical
  Physics}\ }\textbf {\bibinfo {volume} {468}},\ \bibinfo {pages} {9} (\bibinfo
  {year} {2016})}\BibitemShut {NoStop}%
\bibitem [{\citenamefont {Lanzillo}\ \emph {et~al.}(2016)\citenamefont
  {Lanzillo}, \citenamefont {O¡¯Regan},\ and\ \citenamefont
  {Nayak}}]{Lanzillo2016Band}%
  \BibitemOpen
  \bibfield  {author} {\bibinfo {author} {\bibfnamefont {N.~A.}\ \bibnamefont
  {Lanzillo}}, \bibinfo {author} {\bibfnamefont {T.~P.}\ \bibnamefont
  {O¡¯Regan}}, \ and\ \bibinfo {author} {\bibfnamefont {S.~K.}\ \bibnamefont
  {Nayak}},\ }\href@noop {} {\bibfield  {journal} {\bibinfo  {journal}
  {Computational Materials Science}\ }\textbf {\bibinfo {volume} {112}},\
  \bibinfo {pages} {377} (\bibinfo {year} {2016})}\BibitemShut {NoStop}%
\bibitem [{\citenamefont {Chang}(2015)}]{Chang2015Modeling}%
  \BibitemOpen
  \bibfield  {author} {\bibinfo {author} {\bibfnamefont {J.}~\bibnamefont
  {Chang}},\ }\href@noop {} {\bibfield  {journal} {\bibinfo  {journal} {Journal
  of Applied Physics}\ }\textbf {\bibinfo {volume} {117}},\ \bibinfo {pages}
  {214502} (\bibinfo {year} {2015})}\BibitemShut {NoStop}%
\bibitem [{\citenamefont {Chang}\ \emph {et~al.}(2013)\citenamefont {Chang},
  \citenamefont {Register},\ and\ \citenamefont
  {Banerjee}}]{Chang2013Atomistic}%
  \BibitemOpen
  \bibfield  {author} {\bibinfo {author} {\bibfnamefont {J.}~\bibnamefont
  {Chang}}, \bibinfo {author} {\bibfnamefont {L.~F.}\ \bibnamefont {Register}},
  \ and\ \bibinfo {author} {\bibfnamefont {S.~K.}\ \bibnamefont {Banerjee}},\
  }\href@noop {} {\bibfield  {journal} {\bibinfo  {journal} {Applied Physics
  Letters}\ }\textbf {\bibinfo {volume} {103}},\ \bibinfo {pages} {223509 }
  (\bibinfo {year} {2013})}\BibitemShut {NoStop}%
\bibitem [{\citenamefont {Banerjee}\ \emph {et~al.}(2016)\citenamefont
  {Banerjee}, \citenamefont {Mukhopadhyay},\ and\ \citenamefont
  {Sengupta}}]{Banerjee2016Performance}%
  \BibitemOpen
  \bibfield  {author} {\bibinfo {author} {\bibfnamefont {L.}~\bibnamefont
  {Banerjee}}, \bibinfo {author} {\bibfnamefont {A.}~\bibnamefont
  {Mukhopadhyay}}, \ and\ \bibinfo {author} {\bibfnamefont {A.}~\bibnamefont
  {Sengupta}},\ }\href@noop {} {\bibfield  {journal} {\bibinfo  {journal}
  {Journal of Computational Electronics}\ }\textbf {\bibinfo {volume} {15}},\
  \bibinfo {pages} {919} (\bibinfo {year} {2016})}\BibitemShut {NoStop}%
\bibitem [{\citenamefont {Nishiguchi}\ \emph {et~al.}(2015)\citenamefont
  {Nishiguchi}, \citenamefont {Castellanos-Gomez}, \citenamefont {Yamaguchi},
  \citenamefont {Fujiwara}, \citenamefont {Zant},\ and\ \citenamefont
  {Steele}}]{Nishiguchi2015Observing}%
  \BibitemOpen
  \bibfield  {author} {\bibinfo {author} {\bibfnamefont {K.}~\bibnamefont
  {Nishiguchi}}, \bibinfo {author} {\bibfnamefont {A.}~\bibnamefont
  {Castellanos-Gomez}}, \bibinfo {author} {\bibfnamefont {H.}~\bibnamefont
  {Yamaguchi}}, \bibinfo {author} {\bibfnamefont {A.}~\bibnamefont {Fujiwara}},
  \bibinfo {author} {\bibfnamefont {H.~S. J. V.~D.}\ \bibnamefont {Zant}}, \
  and\ \bibinfo {author} {\bibfnamefont {G.~A.}\ \bibnamefont {Steele}},\
  }\href@noop {} {\bibfield  {journal} {\bibinfo  {journal} {Applied Physics
  Letters}\ }\textbf {\bibinfo {volume} {107}},\ \bibinfo {pages} {053101}
  (\bibinfo {year} {2015})}\BibitemShut {NoStop}%
\bibitem [{\citenamefont {Kang}\ \emph {et~al.}(2013)\citenamefont {Kang},
  \citenamefont {Tongay}, \citenamefont {Zhou}, \citenamefont {Li},\ and\
  \citenamefont {Wu}}]{Kang2013Band}%
  \BibitemOpen
  \bibfield  {author} {\bibinfo {author} {\bibfnamefont {J.}~\bibnamefont
  {Kang}}, \bibinfo {author} {\bibfnamefont {S.}~\bibnamefont {Tongay}},
  \bibinfo {author} {\bibfnamefont {J.}~\bibnamefont {Zhou}}, \bibinfo {author}
  {\bibfnamefont {J.}~\bibnamefont {Li}}, \ and\ \bibinfo {author}
  {\bibfnamefont {J.}~\bibnamefont {Wu}},\ }\href@noop {} {\bibfield  {journal}
  {\bibinfo  {journal} {Applied Physics Letters}\ }\textbf {\bibinfo {volume}
  {102}},\ \bibinfo {pages} {012111 } (\bibinfo {year} {2013})}\BibitemShut
  {NoStop}%
\bibitem [{\citenamefont {Han}\ \emph {et~al.}(2011)\citenamefont {Han},
  \citenamefont {Kwon}, \citenamefont {Kim}, \citenamefont {Ryu}, \citenamefont
  {Yun}, \citenamefont {Kim}, \citenamefont {Hwang}, \citenamefont {Kang},
  \citenamefont {Baik}, \citenamefont {Shin},\ and\ \citenamefont
  {Hong}}]{Han2010Band}%
  \BibitemOpen
  \bibfield  {author} {\bibinfo {author} {\bibfnamefont {S.~W.}\ \bibnamefont
  {Han}}, \bibinfo {author} {\bibfnamefont {H.}~\bibnamefont {Kwon}}, \bibinfo
  {author} {\bibfnamefont {S.~K.}\ \bibnamefont {Kim}}, \bibinfo {author}
  {\bibfnamefont {S.}~\bibnamefont {Ryu}}, \bibinfo {author} {\bibfnamefont
  {W.~S.}\ \bibnamefont {Yun}}, \bibinfo {author} {\bibfnamefont {D.~H.}\
  \bibnamefont {Kim}}, \bibinfo {author} {\bibfnamefont {J.~H.}\ \bibnamefont
  {Hwang}}, \bibinfo {author} {\bibfnamefont {J.~S.}\ \bibnamefont {Kang}},
  \bibinfo {author} {\bibfnamefont {J.}~\bibnamefont {Baik}}, \bibinfo {author}
  {\bibfnamefont {H.~J.}\ \bibnamefont {Shin}}, \ and\ \bibinfo {author}
  {\bibfnamefont {S.~C.}\ \bibnamefont {Hong}},\ }\href@noop {} {\bibfield
  {journal} {\bibinfo  {journal} {Physical Review B}\ }\textbf {\bibinfo
  {volume} {84}},\ \bibinfo {pages} {045409} (\bibinfo {year}
  {2011})}\BibitemShut {NoStop}%
\bibitem [{\citenamefont {Das}\ \emph {et~al.}(2012)\citenamefont {Das},
  \citenamefont {Chen}, \citenamefont {Penumatcha},\ and\ \citenamefont
  {Appenzeller}}]{Das2012High}%
  \BibitemOpen
  \bibfield  {author} {\bibinfo {author} {\bibfnamefont {S.}~\bibnamefont
  {Das}}, \bibinfo {author} {\bibfnamefont {H.~Y.}\ \bibnamefont {Chen}},
  \bibinfo {author} {\bibfnamefont {A.~V.}\ \bibnamefont {Penumatcha}}, \ and\
  \bibinfo {author} {\bibfnamefont {J.}~\bibnamefont {Appenzeller}},\
  }\href@noop {} {\bibfield  {journal} {\bibinfo  {journal} {Nano Letters}\
  }\textbf {\bibinfo {volume} {13}},\ \bibinfo {pages} {100} (\bibinfo {year}
  {2012})}\BibitemShut {NoStop}%
\bibitem [{\citenamefont {Singh}\ \emph {et~al.}(2015)\citenamefont {Singh},
  \citenamefont {Hennig}, \citenamefont {Davydov},\ and\ \citenamefont
  {Tavazza}}]{Singh2015Al2O3}%
  \BibitemOpen
  \bibfield  {author} {\bibinfo {author} {\bibfnamefont {A.~K.}\ \bibnamefont
  {Singh}}, \bibinfo {author} {\bibfnamefont {R.~G.}\ \bibnamefont {Hennig}},
  \bibinfo {author} {\bibfnamefont {A.~V.}\ \bibnamefont {Davydov}}, \ and\
  \bibinfo {author} {\bibfnamefont {F.}~\bibnamefont {Tavazza}},\ }\href@noop
  {} {\bibfield  {journal} {\bibinfo  {journal} {Applied Physics Letters}\
  }\textbf {\bibinfo {volume} {107}},\ \bibinfo {pages} {053106} (\bibinfo
  {year} {2015})}\BibitemShut {NoStop}%
\bibitem [{\citenamefont {Son}\ \emph {et~al.}(2015)\citenamefont {Son},
  \citenamefont {Yu}, \citenamefont {Choi}, \citenamefont {Kim},\ and\
  \citenamefont {Choi}}]{Son2015Improved}%
  \BibitemOpen
  \bibfield  {author} {\bibinfo {author} {\bibfnamefont {S.}~\bibnamefont
  {Son}}, \bibinfo {author} {\bibfnamefont {S.}~\bibnamefont {Yu}}, \bibinfo
  {author} {\bibfnamefont {M.}~\bibnamefont {Choi}}, \bibinfo {author}
  {\bibfnamefont {D.}~\bibnamefont {Kim}}, \ and\ \bibinfo {author}
  {\bibfnamefont {C.}~\bibnamefont {Choi}},\ }\href@noop {} {\bibfield
  {journal} {\bibinfo  {journal} {Applied Physics Letters}\ }\textbf {\bibinfo
  {volume} {106}},\ \bibinfo {pages} {021601} (\bibinfo {year}
  {2015})}\BibitemShut {NoStop}%
\bibitem [{\citenamefont {Perdew}\ \emph {et~al.}(1996)\citenamefont {Perdew},
  \citenamefont {Burke},\ and\ \citenamefont
  {Ernzerhof}}]{perdew1996generalized}%
  \BibitemOpen
  \bibfield  {author} {\bibinfo {author} {\bibfnamefont {J.~P.}\ \bibnamefont
  {Perdew}}, \bibinfo {author} {\bibfnamefont {K.}~\bibnamefont {Burke}}, \
  and\ \bibinfo {author} {\bibfnamefont {M.}~\bibnamefont {Ernzerhof}},\
  }\href@noop {} {\bibfield  {journal} {\bibinfo  {journal} {Physical review
  letters}\ }\textbf {\bibinfo {volume} {77}},\ \bibinfo {pages} {3865}
  (\bibinfo {year} {1996})}\BibitemShut {NoStop}%
\bibitem [{\citenamefont {Segall}\ \emph {et~al.}(2002)\citenamefont {Segall},
  \citenamefont {Lindan}, \citenamefont {Probert}, \citenamefont {Pickard},
  \citenamefont {Hasnip}, \citenamefont {Clark},\ and\ \citenamefont
  {Payne}}]{segall2002first-principles}%
  \BibitemOpen
  \bibfield  {author} {\bibinfo {author} {\bibfnamefont {M.~D.}\ \bibnamefont
  {Segall}}, \bibinfo {author} {\bibfnamefont {P.~J.~D.}\ \bibnamefont
  {Lindan}}, \bibinfo {author} {\bibfnamefont {M.}~\bibnamefont {Probert}},
  \bibinfo {author} {\bibfnamefont {C.~J.}\ \bibnamefont {Pickard}}, \bibinfo
  {author} {\bibfnamefont {P.~J.}\ \bibnamefont {Hasnip}}, \bibinfo {author}
  {\bibfnamefont {S.~J.}\ \bibnamefont {Clark}}, \ and\ \bibinfo {author}
  {\bibfnamefont {M.~C.}\ \bibnamefont {Payne}},\ }\href@noop {} {\bibfield
  {journal} {\bibinfo  {journal} {Journal of Physics: Condensed Matter}\
  }\textbf {\bibinfo {volume} {14}},\ \bibinfo {pages} {2717} (\bibinfo {year}
  {2002})}\BibitemShut {NoStop}%
\bibitem [{\citenamefont {Lee}\ \emph {et~al.}(1995)\citenamefont {Lee},
  \citenamefont {Cahill},\ and\ \citenamefont {Allen}}]{Lee1995Thermal}%
  \BibitemOpen
  \bibfield  {author} {\bibinfo {author} {\bibfnamefont {S.}~\bibnamefont
  {Lee}}, \bibinfo {author} {\bibfnamefont {D.~G.}\ \bibnamefont {Cahill}}, \
  and\ \bibinfo {author} {\bibfnamefont {T.~H.}\ \bibnamefont {Allen}},\
  }\href@noop {} {\bibfield  {journal} {\bibinfo  {journal} {Physical Review B
  Condensed Matter}\ }\textbf {\bibinfo {volume} {52}},\ \bibinfo {pages} {253}
  (\bibinfo {year} {1995})}\BibitemShut {NoStop}%
\bibitem [{\citenamefont {Oka}\ \emph {et~al.}(1979)\citenamefont {Oka},
  \citenamefont {Takahashi}, \citenamefont {Okada},\ and\ \citenamefont
  {Iwai}}]{Oka1979Structural}%
  \BibitemOpen
  \bibfield  {author} {\bibinfo {author} {\bibfnamefont {Y.}~\bibnamefont
  {Oka}}, \bibinfo {author} {\bibfnamefont {T.}~\bibnamefont {Takahashi}},
  \bibinfo {author} {\bibfnamefont {K.}~\bibnamefont {Okada}}, \ and\ \bibinfo
  {author} {\bibfnamefont {S.~I.}\ \bibnamefont {Iwai}},\ }\href@noop {}
  {\bibfield  {journal} {\bibinfo  {journal} {Journal of Non-Crystalline
  Solids}\ }\textbf {\bibinfo {volume} {30}},\ \bibinfo {pages} {349} (\bibinfo
  {year} {1979})}\BibitemShut {NoStop}%
\bibitem [{\citenamefont {Heyd}\ \emph {et~al.}(2003)\citenamefont {Heyd},
  \citenamefont {Scuseria},\ and\ \citenamefont {Ernzerhof}}]{heyd2003hybrid}%
  \BibitemOpen
  \bibfield  {author} {\bibinfo {author} {\bibfnamefont {J.}~\bibnamefont
  {Heyd}}, \bibinfo {author} {\bibfnamefont {G.~E.}\ \bibnamefont {Scuseria}},
  \ and\ \bibinfo {author} {\bibfnamefont {M.}~\bibnamefont {Ernzerhof}},\
  }\href@noop {} {\bibfield  {journal} {\bibinfo  {journal} {The Journal of
  Chemical Physics}\ }\textbf {\bibinfo {volume} {118}},\ \bibinfo {pages}
  {8207} (\bibinfo {year} {2003})}\BibitemShut {NoStop}%
\bibitem [{\citenamefont {Heyd}\ and\ \citenamefont
  {Scuseria}(2004)}]{heyd2004efficient}%
  \BibitemOpen
  \bibfield  {author} {\bibinfo {author} {\bibfnamefont {J.}~\bibnamefont
  {Heyd}}\ and\ \bibinfo {author} {\bibfnamefont {G.~E.}\ \bibnamefont
  {Scuseria}},\ }\href@noop {} {\bibfield  {journal} {\bibinfo  {journal} {The
  Journal of chemical physics}\ }\textbf {\bibinfo {volume} {121}},\ \bibinfo
  {pages} {1187} (\bibinfo {year} {2004})}\BibitemShut {NoStop}%
\bibitem [{\citenamefont {Paier}\ \emph {et~al.}(2006)\citenamefont {Paier},
  \citenamefont {Marsman}, \citenamefont {Hummer}, \citenamefont {Kresse},
  \citenamefont {Gerber},\ and\ \citenamefont
  {{\'A}ngy{\'a}n}}]{paier2006erratum}%
  \BibitemOpen
  \bibfield  {author} {\bibinfo {author} {\bibfnamefont {J.}~\bibnamefont
  {Paier}}, \bibinfo {author} {\bibfnamefont {M.}~\bibnamefont {Marsman}},
  \bibinfo {author} {\bibfnamefont {K.}~\bibnamefont {Hummer}}, \bibinfo
  {author} {\bibfnamefont {G.}~\bibnamefont {Kresse}}, \bibinfo {author}
  {\bibfnamefont {I.}~\bibnamefont {Gerber}}, \ and\ \bibinfo {author}
  {\bibfnamefont {J.}~\bibnamefont {{\'A}ngy{\'a}n}},\ }\href@noop {}
  {\bibfield  {journal} {\bibinfo  {journal} {Journal of Chemical Physics}\
  }\textbf {\bibinfo {volume} {125}},\ \bibinfo {pages} {249901} (\bibinfo
  {year} {2006})}\BibitemShut {NoStop}%
\bibitem [{\citenamefont {Heyd}\ \emph {et~al.}(2005)\citenamefont {Heyd},
  \citenamefont {Peralta}, \citenamefont {Scuseria},\ and\ \citenamefont
  {Martin}}]{heyd2005energy}%
  \BibitemOpen
  \bibfield  {author} {\bibinfo {author} {\bibfnamefont {J.}~\bibnamefont
  {Heyd}}, \bibinfo {author} {\bibfnamefont {J.~E.}\ \bibnamefont {Peralta}},
  \bibinfo {author} {\bibfnamefont {G.~E.}\ \bibnamefont {Scuseria}}, \ and\
  \bibinfo {author} {\bibfnamefont {R.~L.}\ \bibnamefont {Martin}},\
  }\href@noop {} {\bibfield  {journal} {\bibinfo  {journal} {The Journal of
  chemical physics}\ }\textbf {\bibinfo {volume} {123}},\ \bibinfo {pages}
  {174101} (\bibinfo {year} {2005})}\BibitemShut {NoStop}%
\bibitem [{\citenamefont {Lyons}\ \emph {et~al.}(2011)\citenamefont {Lyons},
  \citenamefont {Janotti},\ and\ \citenamefont {De~Walle}}]{lyons2011the}%
  \BibitemOpen
  \bibfield  {author} {\bibinfo {author} {\bibfnamefont {J.~L.}\ \bibnamefont
  {Lyons}}, \bibinfo {author} {\bibfnamefont {A.}~\bibnamefont {Janotti}}, \
  and\ \bibinfo {author} {\bibfnamefont {C.~G.~V.}\ \bibnamefont {De~Walle}},\
  }\href@noop {} {\bibfield  {journal} {\bibinfo  {journal} {Microelectronic
  Engineering}\ }\textbf {\bibinfo {volume} {88}},\ \bibinfo {pages} {1452}
  (\bibinfo {year} {2011})}\BibitemShut {NoStop}%
\bibitem [{\citenamefont {De~Walle}\ \emph {et~al.}(2013)\citenamefont
  {De~Walle}, \citenamefont {Choi}, \citenamefont {Weber}, \citenamefont
  {Lyons},\ and\ \citenamefont {Janotti}}]{dewalle2013defects}%
  \BibitemOpen
  \bibfield  {author} {\bibinfo {author} {\bibfnamefont {C.~G.~V.}\
  \bibnamefont {De~Walle}}, \bibinfo {author} {\bibfnamefont {M.}~\bibnamefont
  {Choi}}, \bibinfo {author} {\bibfnamefont {J.~R.}\ \bibnamefont {Weber}},
  \bibinfo {author} {\bibfnamefont {J.~L.}\ \bibnamefont {Lyons}}, \ and\
  \bibinfo {author} {\bibfnamefont {A.}~\bibnamefont {Janotti}},\ }\href@noop
  {} {\bibfield  {journal} {\bibinfo  {journal} {Microelectronic Engineering}\
  }\textbf {\bibinfo {volume} {109}},\ \bibinfo {pages} {211} (\bibinfo {year}
  {2013})}\BibitemShut {NoStop}%
\bibitem [{\citenamefont {Choi}\ \emph
  {et~al.}(2013{\natexlab{b}})\citenamefont {Choi}, \citenamefont {Janotti},\
  and\ \citenamefont {De~Walle}}]{choi2013native}%
  \BibitemOpen
  \bibfield  {author} {\bibinfo {author} {\bibfnamefont {M.}~\bibnamefont
  {Choi}}, \bibinfo {author} {\bibfnamefont {A.}~\bibnamefont {Janotti}}, \
  and\ \bibinfo {author} {\bibfnamefont {C.~G.~V.}\ \bibnamefont {De~Walle}},\
  }\href@noop {} {\bibfield  {journal} {\bibinfo  {journal} {Journal of Applied
  Physics}\ }\textbf {\bibinfo {volume} {113}},\ \bibinfo {pages} {044501}
  (\bibinfo {year} {2013}{\natexlab{b}})}\BibitemShut {NoStop}%
\bibitem [{\citenamefont {Liu}\ \emph {et~al.}(2012)\citenamefont {Liu},
  \citenamefont {Kobayashi}, \citenamefont {Ueno}, \citenamefont {Ohta},
  \citenamefont {Fujioka},\ and\ \citenamefont {Oshima}}]{Liu2012Interfacial}%
  \BibitemOpen
  \bibfield  {author} {\bibinfo {author} {\bibfnamefont {J.~W.}\ \bibnamefont
  {Liu}}, \bibinfo {author} {\bibfnamefont {A.}~\bibnamefont {Kobayashi}},
  \bibinfo {author} {\bibfnamefont {K.}~\bibnamefont {Ueno}}, \bibinfo {author}
  {\bibfnamefont {J.}~\bibnamefont {Ohta}}, \bibinfo {author} {\bibfnamefont
  {H.}~\bibnamefont {Fujioka}}, \ and\ \bibinfo {author} {\bibfnamefont
  {M.}~\bibnamefont {Oshima}},\ }\href@noop {} {\bibfield  {journal} {\bibinfo
  {journal} {e-Journal of Surface Science and Nanotechnology}\ }\textbf
  {\bibinfo {volume} {10}},\ \bibinfo {pages} {165} (\bibinfo {year}
  {2012})}\BibitemShut {NoStop}%
\bibitem [{\citenamefont {Afanas¡¯ev}\ \emph {et~al.}(2006)\citenamefont
  {Afanas¡¯ev}, \citenamefont {Shamuilia}, \citenamefont {Stesmans},
  \citenamefont {Dimoulas}, \citenamefont {Panayiotatos}, \citenamefont
  {Sotiropoulos}, \citenamefont {Houssa},\ and\ \citenamefont
  {Brunco}}]{afanas2006electron}%
  \BibitemOpen
  \bibfield  {author} {\bibinfo {author} {\bibfnamefont {V.}~\bibnamefont
  {Afanas¡¯ev}}, \bibinfo {author} {\bibfnamefont {S.}~\bibnamefont
  {Shamuilia}}, \bibinfo {author} {\bibfnamefont {A.}~\bibnamefont {Stesmans}},
  \bibinfo {author} {\bibfnamefont {A.}~\bibnamefont {Dimoulas}}, \bibinfo
  {author} {\bibfnamefont {Y.}~\bibnamefont {Panayiotatos}}, \bibinfo {author}
  {\bibfnamefont {A.}~\bibnamefont {Sotiropoulos}}, \bibinfo {author}
  {\bibfnamefont {M.}~\bibnamefont {Houssa}}, \ and\ \bibinfo {author}
  {\bibfnamefont {D.}~\bibnamefont {Brunco}},\ }\href@noop {} {\bibfield
  {journal} {\bibinfo  {journal} {Applied physics letters}\ }\textbf {\bibinfo
  {volume} {88}},\ \bibinfo {pages} {132111} (\bibinfo {year}
  {2006})}\BibitemShut {NoStop}%
\bibitem [{\citenamefont {Windhorn}\ \emph {et~al.}(1982)\citenamefont
  {Windhorn}, \citenamefont {Cook},\ and\ \citenamefont
  {Stillman}}]{windhorn1982electron}%
  \BibitemOpen
  \bibfield  {author} {\bibinfo {author} {\bibfnamefont {T.}~\bibnamefont
  {Windhorn}}, \bibinfo {author} {\bibfnamefont {L.}~\bibnamefont {Cook}}, \
  and\ \bibinfo {author} {\bibfnamefont {G.}~\bibnamefont {Stillman}},\
  }\href@noop {} {\bibfield  {journal} {\bibinfo  {journal} {Electron Device
  Letters, IEEE}\ }\textbf {\bibinfo {volume} {3}},\ \bibinfo {pages} {18}
  (\bibinfo {year} {1982})}\BibitemShut {NoStop}%
\bibitem [{\citenamefont {Mak}\ \emph {et~al.}(2010)\citenamefont {Mak},
  \citenamefont {Lee}, \citenamefont {Hone}, \citenamefont {Shan},\ and\
  \citenamefont {Heinz}}]{Mak2010Atomically}%
  \BibitemOpen
  \bibfield  {author} {\bibinfo {author} {\bibfnamefont {K.~F.}\ \bibnamefont
  {Mak}}, \bibinfo {author} {\bibfnamefont {C.}~\bibnamefont {Lee}}, \bibinfo
  {author} {\bibfnamefont {J.}~\bibnamefont {Hone}}, \bibinfo {author}
  {\bibfnamefont {J.}~\bibnamefont {Shan}}, \ and\ \bibinfo {author}
  {\bibfnamefont {T.~F.}\ \bibnamefont {Heinz}},\ }\href@noop {} {\bibfield
  {journal} {\bibinfo  {journal} {Physical Review Letters}\ }\textbf {\bibinfo
  {volume} {105}},\ \bibinfo {pages} {136805} (\bibinfo {year}
  {2010})}\BibitemShut {NoStop}%
\bibitem [{\citenamefont {French}\ \emph {et~al.}(1994)\citenamefont {French},
  \citenamefont {Glass}, \citenamefont {Ohuchi}, \citenamefont {Xu},\ and\
  \citenamefont {Ching}}]{french1994experimental}%
  \BibitemOpen
  \bibfield  {author} {\bibinfo {author} {\bibfnamefont {R.~H.}\ \bibnamefont
  {French}}, \bibinfo {author} {\bibfnamefont {S.~J.}\ \bibnamefont {Glass}},
  \bibinfo {author} {\bibfnamefont {F.~S.}\ \bibnamefont {Ohuchi}}, \bibinfo
  {author} {\bibfnamefont {Y.~N.}\ \bibnamefont {Xu}}, \ and\ \bibinfo {author}
  {\bibfnamefont {W.~Y.}\ \bibnamefont {Ching}},\ }\href@noop {} {\bibfield
  {journal} {\bibinfo  {journal} {Physical Review B}\ }\textbf {\bibinfo
  {volume} {49}},\ \bibinfo {pages} {5133} (\bibinfo {year}
  {1994})}\BibitemShut {NoStop}%
\bibitem [{\citenamefont {He}\ \emph {et~al.}(2007)\citenamefont {He},
  \citenamefont {Zhu}, \citenamefont {Liu}, \citenamefont {Fang},\ and\
  \citenamefont {Zhang}}]{he2007optical}%
  \BibitemOpen
  \bibfield  {author} {\bibinfo {author} {\bibfnamefont {G.}~\bibnamefont
  {He}}, \bibinfo {author} {\bibfnamefont {L.}~\bibnamefont {Zhu}}, \bibinfo
  {author} {\bibfnamefont {M.}~\bibnamefont {Liu}}, \bibinfo {author}
  {\bibfnamefont {Q.~F.}\ \bibnamefont {Fang}}, \ and\ \bibinfo {author}
  {\bibfnamefont {L.~D.}\ \bibnamefont {Zhang}},\ }\href@noop {} {\bibfield
  {journal} {\bibinfo  {journal} {Applied Surface Science}\ }\textbf {\bibinfo
  {volume} {253}},\ \bibinfo {pages} {3413} (\bibinfo {year}
  {2007})}\BibitemShut {NoStop}%
\bibitem [{\citenamefont {Ohta}\ \emph {et~al.}(2004)\citenamefont {Ohta},
  \citenamefont {Yamaoka},\ and\ \citenamefont
  {Miyazaki}}]{ohta2004photoelectron}%
  \BibitemOpen
  \bibfield  {author} {\bibinfo {author} {\bibfnamefont {A.}~\bibnamefont
  {Ohta}}, \bibinfo {author} {\bibfnamefont {M.}~\bibnamefont {Yamaoka}}, \
  and\ \bibinfo {author} {\bibfnamefont {S.}~\bibnamefont {Miyazaki}},\
  }\href@noop {} {\bibfield  {journal} {\bibinfo  {journal} {Microelectronic
  Engineering}\ }\textbf {\bibinfo {volume} {72}},\ \bibinfo {pages} {154}
  (\bibinfo {year} {2004})}\BibitemShut {NoStop}%
\bibitem [{\citenamefont {Qiya}\ \emph {et~al.}(2014)\citenamefont {Qiya},
  \citenamefont {Zebo}, \citenamefont {Ting}, \citenamefont {Shiyan},
  \citenamefont {Yongsheng}, \citenamefont {Jiajun},\ and\ \citenamefont
  {Yanyan}}]{qiya2014band}%
  \BibitemOpen
  \bibfield  {author} {\bibinfo {author} {\bibfnamefont {L.}~\bibnamefont
  {Qiya}}, \bibinfo {author} {\bibfnamefont {F.}~\bibnamefont {Zebo}}, \bibinfo
  {author} {\bibfnamefont {J.}~\bibnamefont {Ting}}, \bibinfo {author}
  {\bibfnamefont {L.}~\bibnamefont {Shiyan}}, \bibinfo {author} {\bibfnamefont
  {T.}~\bibnamefont {Yongsheng}}, \bibinfo {author} {\bibfnamefont
  {C.}~\bibnamefont {Jiajun}}, \ and\ \bibinfo {author} {\bibfnamefont
  {Z.}~\bibnamefont {Yanyan}},\ }\href@noop {} {\bibfield  {journal} {\bibinfo
  {journal} {Chinese Physics Letters}\ }\textbf {\bibinfo {volume} {31}},\
  \bibinfo {pages} {027702} (\bibinfo {year} {2014})}\BibitemShut {NoStop}%
\bibitem [{\citenamefont {French}(1990)}]{1990electronic}%
  \BibitemOpen
  \bibfield  {author} {\bibinfo {author} {\bibfnamefont {R.~H.}\ \bibnamefont
  {French}},\ }\href@noop {} {\bibfield  {journal} {\bibinfo  {journal}
  {Journal of the American Ceramic Society}\ }\textbf {\bibinfo {volume}
  {73}},\ \bibinfo {pages} {477¨C489} (\bibinfo {year} {1990})}\BibitemShut
  {NoStop}%
\bibitem [{\citenamefont {Silvestri}\ \emph {et~al.}(2013)\citenamefont
  {Silvestri}, \citenamefont {Cervenka}, \citenamefont {Prawer},\ and\
  \citenamefont {Ladouceur}}]{silvestri2013first}%
  \BibitemOpen
  \bibfield  {author} {\bibinfo {author} {\bibfnamefont {L.}~\bibnamefont
  {Silvestri}}, \bibinfo {author} {\bibfnamefont {J.}~\bibnamefont {Cervenka}},
  \bibinfo {author} {\bibfnamefont {S.}~\bibnamefont {Prawer}}, \ and\ \bibinfo
  {author} {\bibfnamefont {F.}~\bibnamefont {Ladouceur}},\ }\href@noop {}
  {\bibfield  {journal} {\bibinfo  {journal} {Diamond and Related Materials}\
  }\textbf {\bibinfo {volume} {31}},\ \bibinfo {pages} {25} (\bibinfo {year}
  {2013})}\BibitemShut {NoStop}%
\bibitem [{\citenamefont {D$'$Amico}\ \emph {et~al.}(2012)\citenamefont
  {D$'$Amico}, \citenamefont {Cantele},\ and\ \citenamefont
  {Ninno}}]{d2012first}%
  \BibitemOpen
  \bibfield  {author} {\bibinfo {author} {\bibfnamefont {N.~R.}\ \bibnamefont
  {D$'$Amico}}, \bibinfo {author} {\bibfnamefont {G.}~\bibnamefont {Cantele}},
  \ and\ \bibinfo {author} {\bibfnamefont {D.}~\bibnamefont {Ninno}},\
  }\href@noop {} {\bibfield  {journal} {\bibinfo  {journal} {Applied Physics
  Letters}\ }\textbf {\bibinfo {volume} {101}},\ \bibinfo {pages} {141606}
  (\bibinfo {year} {2012})}\BibitemShut {NoStop}%
\bibitem [{\citenamefont {Weber}\ \emph {et~al.}(2011)\citenamefont {Weber},
  \citenamefont {Janotti},\ and\ \citenamefont {Walle}}]{weber2011native}%
  \BibitemOpen
  \bibfield  {author} {\bibinfo {author} {\bibfnamefont {J.~R.}\ \bibnamefont
  {Weber}}, \bibinfo {author} {\bibfnamefont {A.}~\bibnamefont {Janotti}}, \
  and\ \bibinfo {author} {\bibfnamefont {C.~G. V.~D.}\ \bibnamefont {Walle}},\
  }\href@noop {} {\bibfield  {journal} {\bibinfo  {journal} {Journal of Applied
  Physics}\ }\textbf {\bibinfo {volume} {109}},\ \bibinfo {pages} {033715}
  (\bibinfo {year} {2011})}\BibitemShut {NoStop}%
\bibitem [{\citenamefont {Liu}\ \emph {et~al.}(2007)\citenamefont {Liu},
  \citenamefont {Zheng},\ and\ \citenamefont {Jiang}}]{Liu2007First}%
  \BibitemOpen
  \bibfield  {author} {\bibinfo {author} {\bibfnamefont {W.}~\bibnamefont
  {Liu}}, \bibinfo {author} {\bibfnamefont {W.~T.}\ \bibnamefont {Zheng}}, \
  and\ \bibinfo {author} {\bibfnamefont {Q.}~\bibnamefont {Jiang}},\
  }\href@noop {} {\bibfield  {journal} {\bibinfo  {journal} {Physical Review
  B}\ }\textbf {\bibinfo {volume} {75}},\ \bibinfo {pages} {235322} (\bibinfo
  {year} {2007})}\BibitemShut {NoStop}%
\bibitem [{\citenamefont {Shi}\ \emph {et~al.}(2016)\citenamefont {Shi},
  \citenamefont {Liu},\ and\ \citenamefont {Dong}}]{Li2016Investigation}%
  \BibitemOpen
  \bibfield  {author} {\bibinfo {author} {\bibfnamefont {L.-B.}\ \bibnamefont
  {Shi}}, \bibinfo {author} {\bibfnamefont {X.-Y.}\ \bibnamefont {Liu}}, \ and\
  \bibinfo {author} {\bibfnamefont {H.-K.}\ \bibnamefont {Dong}},\ }\href@noop
  {} {\bibfield  {journal} {\bibinfo  {journal} {Journal of Applied Physics}\
  }\textbf {\bibinfo {volume} {120}},\ \bibinfo {pages} {105306} (\bibinfo
  {year} {2016})}\BibitemShut {NoStop}%
\bibitem [{\citenamefont {Yang}\ \emph {et~al.}(2014)\citenamefont {Yang},
  \citenamefont {Fan}, \citenamefont {Liu},\ and\ \citenamefont
  {Ran}}]{Yang2014First}%
  \BibitemOpen
  \bibfield  {author} {\bibinfo {author} {\bibfnamefont {Y.~L.}\ \bibnamefont
  {Yang}}, \bibinfo {author} {\bibfnamefont {X.~L.}\ \bibnamefont {Fan}},
  \bibinfo {author} {\bibfnamefont {C.}~\bibnamefont {Liu}}, \ and\ \bibinfo
  {author} {\bibfnamefont {R.~X.}\ \bibnamefont {Ran}},\ }\href@noop {}
  {\bibfield  {journal} {\bibinfo  {journal} {Physica B}\ }\textbf {\bibinfo
  {volume} {434}},\ \bibinfo {pages} {7} (\bibinfo {year} {2014})}\BibitemShut
  {NoStop}%
\bibitem [{\citenamefont {Yeo}\ \emph {et~al.}(2000)\citenamefont {Yeo},
  \citenamefont {Lu}, \citenamefont {Lee}, \citenamefont {King}, \citenamefont
  {Hu}, \citenamefont {Wang}, \citenamefont {Guo},\ and\ \citenamefont
  {Ma}}]{yeo2000direct}%
  \BibitemOpen
  \bibfield  {author} {\bibinfo {author} {\bibfnamefont {Y.~C.}\ \bibnamefont
  {Yeo}}, \bibinfo {author} {\bibfnamefont {Q.}~\bibnamefont {Lu}}, \bibinfo
  {author} {\bibfnamefont {W.~C.}\ \bibnamefont {Lee}}, \bibinfo {author}
  {\bibfnamefont {T.-J.}\ \bibnamefont {King}}, \bibinfo {author}
  {\bibfnamefont {C.}~\bibnamefont {Hu}}, \bibinfo {author} {\bibfnamefont
  {X.}~\bibnamefont {Wang}}, \bibinfo {author} {\bibfnamefont {X.}~\bibnamefont
  {Guo}}, \ and\ \bibinfo {author} {\bibfnamefont {T.}~\bibnamefont {Ma}},\
  }\href@noop {} {\bibfield  {journal} {\bibinfo  {journal} {Electron Device
  Letters, IEEE}\ }\textbf {\bibinfo {volume} {21}},\ \bibinfo {pages} {540}
  (\bibinfo {year} {2000})}\BibitemShut {NoStop}%
\bibitem [{\citenamefont {Das}\ \emph {et~al.}(2010)\citenamefont {Das},
  \citenamefont {Dalapati}, \citenamefont {Chi}, \citenamefont {Biswas},\ and\
  \citenamefont {Maiti}}]{das2010characterization}%
  \BibitemOpen
  \bibfield  {author} {\bibinfo {author} {\bibfnamefont {P.}~\bibnamefont
  {Das}}, \bibinfo {author} {\bibfnamefont {G.}~\bibnamefont {Dalapati}},
  \bibinfo {author} {\bibfnamefont {D.}~\bibnamefont {Chi}}, \bibinfo {author}
  {\bibfnamefont {A.}~\bibnamefont {Biswas}}, \ and\ \bibinfo {author}
  {\bibfnamefont {C.}~\bibnamefont {Maiti}},\ }\href@noop {} {\bibfield
  {journal} {\bibinfo  {journal} {Applied Surface Science}\ }\textbf {\bibinfo
  {volume} {256}},\ \bibinfo {pages} {2245} (\bibinfo {year}
  {2010})}\BibitemShut {NoStop}%
\bibitem [{\citenamefont {Van~de Walle}\ \emph {et~al.}(2013)\citenamefont
  {Van~de Walle}, \citenamefont {Choi}, \citenamefont {Weber}, \citenamefont
  {Lyons},\ and\ \citenamefont {Janotti}}]{van2013defects}%
  \BibitemOpen
  \bibfield  {author} {\bibinfo {author} {\bibfnamefont {C.}~\bibnamefont
  {Van~de Walle}}, \bibinfo {author} {\bibfnamefont {M.}~\bibnamefont {Choi}},
  \bibinfo {author} {\bibfnamefont {J.}~\bibnamefont {Weber}}, \bibinfo
  {author} {\bibfnamefont {J.}~\bibnamefont {Lyons}}, \ and\ \bibinfo {author}
  {\bibfnamefont {A.}~\bibnamefont {Janotti}},\ }\href@noop {} {\bibfield
  {journal} {\bibinfo  {journal} {Microelectronic Engineering}\ }\textbf
  {\bibinfo {volume} {109}},\ \bibinfo {pages} {211} (\bibinfo {year}
  {2013})}\BibitemShut {NoStop}%
\bibitem [{\citenamefont {Sun}\ \emph {et~al.}(2008{\natexlab{a}})\citenamefont
  {Sun}, \citenamefont {Shi}, \citenamefont {Dong}, \citenamefont {Liu},
  \citenamefont {Ding},\ and\ \citenamefont {Zhang}}]{sun2008impact}%
  \BibitemOpen
  \bibfield  {author} {\bibinfo {author} {\bibfnamefont {Q.~Q.}\ \bibnamefont
  {Sun}}, \bibinfo {author} {\bibfnamefont {Y.}~\bibnamefont {Shi}}, \bibinfo
  {author} {\bibfnamefont {L.}~\bibnamefont {Dong}}, \bibinfo {author}
  {\bibfnamefont {H.}~\bibnamefont {Liu}}, \bibinfo {author} {\bibfnamefont
  {S.~J.}\ \bibnamefont {Ding}}, \ and\ \bibinfo {author} {\bibfnamefont
  {D.~W.}\ \bibnamefont {Zhang}},\ }\href@noop {} {\bibfield  {journal}
  {\bibinfo  {journal} {Applied Physics Letters}\ }\textbf {\bibinfo {volume}
  {92}},\ \bibinfo {pages} {102908} (\bibinfo {year}
  {2008}{\natexlab{a}})}\BibitemShut {NoStop}%
\bibitem [{\citenamefont {Sun}\ \emph {et~al.}(2008{\natexlab{b}})\citenamefont
  {Sun}, \citenamefont {Zhang}, \citenamefont {Dong}, \citenamefont {Shi},
  \citenamefont {Ding},\ and\ \citenamefont {Zhang}}]{sun2008effect}%
  \BibitemOpen
  \bibfield  {author} {\bibinfo {author} {\bibfnamefont {Q.-Q.}\ \bibnamefont
  {Sun}}, \bibinfo {author} {\bibfnamefont {C.}~\bibnamefont {Zhang}}, \bibinfo
  {author} {\bibfnamefont {L.}~\bibnamefont {Dong}}, \bibinfo {author}
  {\bibfnamefont {Y.}~\bibnamefont {Shi}}, \bibinfo {author} {\bibfnamefont
  {S.-J.}\ \bibnamefont {Ding}}, \ and\ \bibinfo {author} {\bibfnamefont
  {D.~W.}\ \bibnamefont {Zhang}},\ }\href@noop {} {\bibfield  {journal}
  {\bibinfo  {journal} {Journal of Applied Physics}\ }\textbf {\bibinfo
  {volume} {103}},\ \bibinfo {pages} {114102} (\bibinfo {year}
  {2008}{\natexlab{b}})}\BibitemShut {NoStop}%
\bibitem [{\citenamefont {Sun}\ \emph {et~al.}(2007)\citenamefont {Sun},
  \citenamefont {Chen}, \citenamefont {Ding}, \citenamefont {Xu}, \citenamefont
  {Lu}, \citenamefont {Lindhrengifo}, \citenamefont {Zhang},\ and\
  \citenamefont {Wang}}]{sun2007comparative}%
  \BibitemOpen
  \bibfield  {author} {\bibinfo {author} {\bibfnamefont {Q.~Q.}\ \bibnamefont
  {Sun}}, \bibinfo {author} {\bibfnamefont {W.}~\bibnamefont {Chen}}, \bibinfo
  {author} {\bibfnamefont {S.~J.}\ \bibnamefont {Ding}}, \bibinfo {author}
  {\bibfnamefont {M.}~\bibnamefont {Xu}}, \bibinfo {author} {\bibfnamefont
  {H.~L.}\ \bibnamefont {Lu}}, \bibinfo {author} {\bibfnamefont {H.~C.}\
  \bibnamefont {Lindhrengifo}}, \bibinfo {author} {\bibfnamefont {D.~W.}\
  \bibnamefont {Zhang}}, \ and\ \bibinfo {author} {\bibfnamefont {L.~K.}\
  \bibnamefont {Wang}},\ }\href@noop {} {\bibfield  {journal} {\bibinfo
  {journal} {Applied Physics Letters}\ }\textbf {\bibinfo {volume} {90}},\
  \bibinfo {pages} {142904} (\bibinfo {year} {2007})}\BibitemShut {NoStop}%
\bibitem [{\citenamefont {Kaloni}\ \emph {et~al.}(2013)\citenamefont {Kaloni},
  \citenamefont {Cheng},\ and\ \citenamefont
  {Schwingenschl{\"o}gl}}]{kaloni2013hole}%
  \BibitemOpen
  \bibfield  {author} {\bibinfo {author} {\bibfnamefont {T.~P.}\ \bibnamefont
  {Kaloni}}, \bibinfo {author} {\bibfnamefont {Y.}~\bibnamefont {Cheng}}, \
  and\ \bibinfo {author} {\bibfnamefont {U.}~\bibnamefont
  {Schwingenschl{\"o}gl}},\ }\href@noop {} {\bibfield  {journal} {\bibinfo
  {journal} {Journal of Applied Physics}\ }\textbf {\bibinfo {volume} {113}},\
  \bibinfo {pages} {104305} (\bibinfo {year} {2013})}\BibitemShut {NoStop}%
\bibitem [{\citenamefont {Liu}\ \emph {et~al.}(2015{\natexlab{a}})\citenamefont
  {Liu}, \citenamefont {Wu}, \citenamefont {Liu},\ and\ \citenamefont
  {Chu}}]{liu2015strain}%
  \BibitemOpen
  \bibfield  {author} {\bibinfo {author} {\bibfnamefont {L.}~\bibnamefont
  {Liu}}, \bibinfo {author} {\bibfnamefont {X.}~\bibnamefont {Wu}}, \bibinfo
  {author} {\bibfnamefont {X.}~\bibnamefont {Liu}}, \ and\ \bibinfo {author}
  {\bibfnamefont {P.~K.}\ \bibnamefont {Chu}},\ }\href@noop {} {\bibfield
  {journal} {\bibinfo  {journal} {Applied Surface Science}\ }\textbf {\bibinfo
  {volume} {356}},\ \bibinfo {pages} {626} (\bibinfo {year}
  {2015}{\natexlab{a}})}\BibitemShut {NoStop}%
\bibitem [{\citenamefont {Yu}\ \emph {et~al.}(2015)\citenamefont {Yu},
  \citenamefont {Xiong}, \citenamefont {Eshun}, \citenamefont {Yuan},\ and\
  \citenamefont {Li}}]{yu2015phase}%
  \BibitemOpen
  \bibfield  {author} {\bibinfo {author} {\bibfnamefont {S.}~\bibnamefont
  {Yu}}, \bibinfo {author} {\bibfnamefont {H.~D.}\ \bibnamefont {Xiong}},
  \bibinfo {author} {\bibfnamefont {K.}~\bibnamefont {Eshun}}, \bibinfo
  {author} {\bibfnamefont {H.}~\bibnamefont {Yuan}}, \ and\ \bibinfo {author}
  {\bibfnamefont {Q.}~\bibnamefont {Li}},\ }\href@noop {} {\bibfield  {journal}
  {\bibinfo  {journal} {Applied Surface Science}\ }\textbf {\bibinfo {volume}
  {325}},\ \bibinfo {pages} {27} (\bibinfo {year} {2015})}\BibitemShut
  {NoStop}%
\bibitem [{\citenamefont {Chang}\ \emph {et~al.}(2015)\citenamefont {Chang},
  \citenamefont {Liu}, \citenamefont {Zeng},\ and\ \citenamefont
  {Du}}]{chang2015hole}%
  \BibitemOpen
  \bibfield  {author} {\bibinfo {author} {\bibfnamefont {P.}~\bibnamefont
  {Chang}}, \bibinfo {author} {\bibfnamefont {X.}~\bibnamefont {Liu}}, \bibinfo
  {author} {\bibfnamefont {L.}~\bibnamefont {Zeng}}, \ and\ \bibinfo {author}
  {\bibfnamefont {G.}~\bibnamefont {Du}},\ }\href@noop {} {\bibfield  {journal}
  {\bibinfo  {journal} {Solid-State Electronics}\ }\textbf {\bibinfo {volume}
  {113}},\ \bibinfo {pages} {68} (\bibinfo {year} {2015})}\BibitemShut
  {NoStop}%
\bibitem [{\citenamefont {Zhang}\ \emph {et~al.}(2009)\citenamefont {Zhang},
  \citenamefont {Fischetti}, \citenamefont {Sor{\'e}e}, \citenamefont {Magnus},
  \citenamefont {Heyns},\ and\ \citenamefont {Meuris}}]{zhang2009physical}%
  \BibitemOpen
  \bibfield  {author} {\bibinfo {author} {\bibfnamefont {Y.}~\bibnamefont
  {Zhang}}, \bibinfo {author} {\bibfnamefont {M.}~\bibnamefont {Fischetti}},
  \bibinfo {author} {\bibfnamefont {B.}~\bibnamefont {Sor{\'e}e}}, \bibinfo
  {author} {\bibfnamefont {W.}~\bibnamefont {Magnus}}, \bibinfo {author}
  {\bibfnamefont {M.}~\bibnamefont {Heyns}}, \ and\ \bibinfo {author}
  {\bibfnamefont {M.}~\bibnamefont {Meuris}},\ }\href@noop {} {\bibfield
  {journal} {\bibinfo  {journal} {Journal of applied physics}\ }\textbf
  {\bibinfo {volume} {106}},\ \bibinfo {pages} {083704} (\bibinfo {year}
  {2009})}\BibitemShut {NoStop}%
\bibitem [{\citenamefont {Fiori}\ \emph {et~al.}(2013)\citenamefont {Fiori},
  \citenamefont {Gallois-Garreignot}, \citenamefont {Jaouen},\ and\
  \citenamefont {Tavernier}}]{fiori2013strain}%
  \BibitemOpen
  \bibfield  {author} {\bibinfo {author} {\bibfnamefont {V.}~\bibnamefont
  {Fiori}}, \bibinfo {author} {\bibfnamefont {S.}~\bibnamefont
  {Gallois-Garreignot}}, \bibinfo {author} {\bibfnamefont {H.}~\bibnamefont
  {Jaouen}}, \ and\ \bibinfo {author} {\bibfnamefont {C.}~\bibnamefont
  {Tavernier}},\ }\href@noop {} {\bibfield  {journal} {\bibinfo  {journal}
  {Microelectronics Reliability}\ }\textbf {\bibinfo {volume} {53}},\ \bibinfo
  {pages} {229} (\bibinfo {year} {2013})}\BibitemShut {NoStop}%
\bibitem [{\citenamefont {Cheng}\ \emph {et~al.}(2013)\citenamefont {Cheng},
  \citenamefont {Lee}, \citenamefont {Chang}, \citenamefont {Lin},
  \citenamefont {Chen},\ and\ \citenamefont {Hsieh}}]{cheng2013uniaxial}%
  \BibitemOpen
  \bibfield  {author} {\bibinfo {author} {\bibfnamefont {S.-Y.}\ \bibnamefont
  {Cheng}}, \bibinfo {author} {\bibfnamefont {M.-H.}\ \bibnamefont {Lee}},
  \bibinfo {author} {\bibfnamefont {S.}~\bibnamefont {Chang}}, \bibinfo
  {author} {\bibfnamefont {C.-Y.}\ \bibnamefont {Lin}}, \bibinfo {author}
  {\bibfnamefont {K.-T.}\ \bibnamefont {Chen}}, \ and\ \bibinfo {author}
  {\bibfnamefont {B.-F.}\ \bibnamefont {Hsieh}},\ }\href@noop {} {\bibfield
  {journal} {\bibinfo  {journal} {Thin Solid Films}\ }\textbf {\bibinfo
  {volume} {544}},\ \bibinfo {pages} {487} (\bibinfo {year}
  {2013})}\BibitemShut {NoStop}%
\bibitem [{\citenamefont {Chern}\ \emph {et~al.}(2014)\citenamefont {Chern},
  \citenamefont {Hashemi}, \citenamefont {Teherani}, \citenamefont
  {Antoniadis},\ and\ \citenamefont {Hoyt}}]{chern2014record}%
  \BibitemOpen
  \bibfield  {author} {\bibinfo {author} {\bibfnamefont {W.-T.}\ \bibnamefont
  {Chern}}, \bibinfo {author} {\bibfnamefont {P.}~\bibnamefont {Hashemi}},
  \bibinfo {author} {\bibfnamefont {J.~T.}\ \bibnamefont {Teherani}}, \bibinfo
  {author} {\bibfnamefont {D.~A.}\ \bibnamefont {Antoniadis}}, \ and\ \bibinfo
  {author} {\bibfnamefont {J.~L.}\ \bibnamefont {Hoyt}},\ }\href@noop {}
  {\bibfield  {journal} {\bibinfo  {journal} {Electron Device Letters, IEEE}\
  }\textbf {\bibinfo {volume} {35}},\ \bibinfo {pages} {309} (\bibinfo {year}
  {2014})}\BibitemShut {NoStop}%
\bibitem [{\citenamefont {Shi}\ \emph {et~al.}(2017)\citenamefont {Shi},
  \citenamefont {Li}, \citenamefont {Xiu}, \citenamefont {Liu}, \citenamefont
  {Zhang}, \citenamefont {Li},\ and\ \citenamefont {Dong}}]{Shi2017First}%
  \BibitemOpen
  \bibfield  {author} {\bibinfo {author} {\bibfnamefont {L.~B.}\ \bibnamefont
  {Shi}}, \bibinfo {author} {\bibfnamefont {M.~B.}\ \bibnamefont {Li}},
  \bibinfo {author} {\bibfnamefont {X.~M.}\ \bibnamefont {Xiu}}, \bibinfo
  {author} {\bibfnamefont {X.~Y.}\ \bibnamefont {Liu}}, \bibinfo {author}
  {\bibfnamefont {K.~C.}\ \bibnamefont {Zhang}}, \bibinfo {author}
  {\bibfnamefont {C.~R.}\ \bibnamefont {Li}}, \ and\ \bibinfo {author}
  {\bibfnamefont {H.~K.}\ \bibnamefont {Dong}},\ }\href@noop {} {\bibfield
  {journal} {\bibinfo  {journal} {Physica B}\ }\textbf {\bibinfo {volume}
  {510}},\ \bibinfo {pages} {13} (\bibinfo {year} {2017})}\BibitemShut
  {NoStop}%
\bibitem [{\citenamefont {Liu}\ \emph {et~al.}(2015{\natexlab{b}})\citenamefont
  {Liu}, \citenamefont {Xiao}, \citenamefont {Yao}, \citenamefont {Xu},\ and\
  \citenamefont {Yao}}]{Liu2015Electronic}%
  \BibitemOpen
  \bibfield  {author} {\bibinfo {author} {\bibfnamefont {G.~B.}\ \bibnamefont
  {Liu}}, \bibinfo {author} {\bibfnamefont {D.}~\bibnamefont {Xiao}}, \bibinfo
  {author} {\bibfnamefont {Y.}~\bibnamefont {Yao}}, \bibinfo {author}
  {\bibfnamefont {X.}~\bibnamefont {Xu}}, \ and\ \bibinfo {author}
  {\bibfnamefont {W.}~\bibnamefont {Yao}},\ }\href@noop {} {\bibfield
  {journal} {\bibinfo  {journal} {Chemical Society Reviews}\ }\textbf {\bibinfo
  {volume} {44}},\ \bibinfo {pages} {2643} (\bibinfo {year}
  {2015}{\natexlab{b}})}\BibitemShut {NoStop}%
\bibitem [{\citenamefont {Mohammad~Tabatabaei}\ \emph
  {et~al.}(2013)\citenamefont {Mohammad~Tabatabaei}, \citenamefont {Noei},
  \citenamefont {Khaliji},\ and\ \citenamefont {Pourfath}}]{mohammad2013a}%
  \BibitemOpen
  \bibfield  {author} {\bibinfo {author} {\bibfnamefont {S.}~\bibnamefont
  {Mohammad~Tabatabaei}}, \bibinfo {author} {\bibfnamefont {M.}~\bibnamefont
  {Noei}}, \bibinfo {author} {\bibfnamefont {K.}~\bibnamefont {Khaliji}}, \
  and\ \bibinfo {author} {\bibfnamefont {M.}~\bibnamefont {Pourfath}},\
  }\href@noop {} {\bibfield  {journal} {\bibinfo  {journal} {Journal of Applied
  Physics}\ }\textbf {\bibinfo {volume} {113}},\ \bibinfo {pages} {163708}
  (\bibinfo {year} {2013})}\BibitemShut {NoStop}%
\end{thebibliography}

%

\end{document}